\documentclass[aip,pop,preprint]{revtex4-1}
\usepackage{graphicx}
\usepackage{amsmath}
\usepackage{amssymb}
\usepackage{bm}
\usepackage{natbib}
\usepackage{color}
\usepackage[normalem]{ulem}

\newcommand{\grad}{\boldsymbol{\nabla}}
\newcommand{\cross}{\boldsymbol{\times}}

\newcommand{\ue}{\ensuremath{\mathbf{u}_{_E}}}

\newcommand{\ave}[1]{\ensuremath{\left<#1\right>}}
\newcommand{\uek}{\ensuremath{\mathbf{u_E} \boldsymbol{\cdot} \boldsymbol{\kappa}}}

\newcommand{\myemail}{joel.t.dahlin@nasa.gov}

\begin{document}

\preprint{AIP/123-QED}


\title{The role of three-dimensional transport in driving enhanced electron acceleration during magnetic reconnection}

\author{J. T. Dahlin}
\altaffiliation{University Corporation for Atmospheric Research, Boulder, CO, USA}
\altaffiliation{NASA Goddard Space Flight Center, Greenbelt, Maryland 20771, USA}
\altaffiliation{Jack Eddy Postdoctoral Fellow}
\email[]{\myemail}
\author{J. F. Drake}
\altaffiliation{Department of Physics, University of Maryland, College Park, Maryland 20742, USA}
\altaffiliation{Institute for Physical Science and Technology, University of Maryland, College Park, Maryland 20742, USA}

\author{M. Swisdak}

\affiliation{Institute for Research in Electronics and Applied Physics, 
University of Maryland, College Park, Maryland 20742, USA}

\begin{abstract}
Magnetic reconnection is an important driver of energetic particles in
many astrophysical phenomena. Using kinetic particle-in-cell (PIC) simulations, we 
explore the impact of three-dimensional reconnection dynamics on the 
efficiency of particle acceleration. In two-dimensional systems, Alfv\'enic outflows 
expel energetic electrons into flux ropes where they become trapped and disconnected from acceleration
regions. However, in three-dimensional systems these flux ropes develop axial structure that
enables particles to leak out and return to acceleration regions. This requires a finite 
guide field so that particles may move quickly along the flux rope axis. We show that greatest
energetic electron production occurs when the guide field is of the same order as the reconnecting
component: large enough to facilitate strong transport, but not so large as to throttle the dominant Fermi 
mechanism responsible for efficient electron acceleration.
This suggests a natural explanation for the envelope of 
electron acceleration during the impulsive phase of eruptive flares.

\end{abstract}


\pacs{52.35.Vd,94.30.cp,52.65.Rr,96.60.Iv}

\maketitle


\section{Introduction}
Magnetic reconnection is thought to be an important driver of energetic particles in 
astrophysical plasmas, releasing stored magnetic energy via efficient acceleration 
of a non-thermal population. Reconnection-associated energetic particle production has
been well-observed in solar flares \cite{lin03a} and magnetospheric storms
\citep{oieroset02a}. Solar flare observations in particular indicate that reconnection-driven
acceleration can be very efficient, driving a large non-thermal electron component
with a total energy content comparable to that of the energy in the initial magnetic field 
\cite{krucker10a,oka13a}. This mechanism therefore provides a promising explanation
for a variety of astrophysical phenomena characterized by energetic particle 
production, including stellar flares, gamma-ray bursts \cite{drenkhahn02a,michel94a}, 
and gamma-ray flares in pulsar wind nebulae \cite{tavani11a,abdo11a}. 

Electron acceleration by magnetic reconnection has attracted
significant interest, e.g. 
\cite{hoshino01a,zenitani01a,drake05a,pritchett06a,egedal09a,oka10a,hoshino12a}.
Two specific processes have received the most attention. 
The first is acceleration by electric fields parallel to the local magnetic field
($E_\parallel$) \citep{litvinenko96a,drake05a,uzdensky11a,egedal12a}. 
However, the number of electrons that can be accelerated through this mechanism 
can be limited because during magnetic reconnection non-zero $E_{\parallel}$ typically 
only occur near X-lines and separatrices. Additionally, acceleration by parallel
electric fields has a weak energy-scaling \cite{dahlin15a} ($\sim \epsilon^{1/2}$
with $\epsilon$ the particle energy) and characteristically
drives bulk electron heating rather than a non-thermal component.

In the second process \citep{drake06a}, charged particles gain energy
as they reflect from the ends of contracting magnetic islands.  (An
analogous process occurs during the acceleration of cosmic rays by the
first-order Fermi mechanism.) In contrast to the localization of
$E_{\parallel}$, this can occur wherever there are contracting field
lines, including the merging of magnetic islands and in the outflows of
single X-line reconnection
\citep{drake06a,oka10a,drake10a,drake13a,hoshino12a} and in turbulent
reconnecting systems where magnetic field lines are stochastic and
conventional islands do not exist \cite{dahlin15a}.  This mechanism is
therefore volume-filling and can accelerate a large number of
particles. This mechanism scales strongly with the particle energy ($\sim \epsilon$) and
preferentially energizes non-thermal particles \cite{dahlin15a,dahlin16a}.

Several recent studies of two-dimensional reconnection
\cite{dahlin14a,wang16a,dahlin16a} found that the
guide field (the magnetic component parallel to the reconnection axis)
controls which mechanisms contribute to electron energy gain.
In the antiparallel (small guide field) regime Fermi reflection dominates
\cite{dahlin14a,dahlin16a,guo14a,li15a,wang16a}, whereas in reconnection
where the guide field is much larger than the reconnecting component 
parallel electric fields drive essentially all of the electron energy gain
\cite{numata15a,dahlin16a}. In the latter (strong guide field) regime
energetic electron production is weak, indicating that parallel electric
fields are ineffective drivers of energetic electrons in reconnection \cite{dahlin16a}.

Studies of particle acceleration in reconnection have primarily been based
on 2D simulations,
in which accelerated particles are typically localized near the X-line, along
magnetic separatrices and within magnetic islands \citep{drake05a,guo14a}. There
are some observations with small ambient guide fields \citep{chen08a,retino08a,huang12a}
that support such a picture. A notable exception are Wind observations
in which energetic electrons up to 300 keV are seen for more than an hour in an
extended region around the reconnection region \citep{oieroset02a}. These observations
correspond to reconnection with a strong guide field.

Two-dimensional simulations impose limitations on the magnetic topology as well as
the available spectrum of instabilities. In the presence of an ambient guide
field, 3D reconnection can become turbulent as a result of the generation of 
magnetic islands along separatrices and adjacent surfaces \cite{schreier10a,daughton11a}. 
While test particle trajectories in MHD fields have been used to explore acceleration
in such systems\citep{onofri06a,kowal11a}, the absence of feedback of energetic
particles on the reconnection process in such models limits their applicability
to physical systems. Recent 3D studies of kinetic reconnection examined particle
acceleration in pair plasmas \citep{sironi14a,guo14a}.
However, these studies focused on relativistic regimes where the magnetic energy
per particle exceeds the rest mass energy and included no ambient guide field.

In a recent kinetic study of nonrelativistic reconnection, we showed that 
energetic electron production was greatly enhanced in three-dimensional systems \cite{dahlin15a}.
This occurs because two-dimensional magnetic islands trap particles, limiting 
energy gain, whereas three-dimensional reconnection generates a stochastic field 
that enables electrons to access volume-filling acceleration regions. The relative 
enhancement was found to increase with the size of the simulation domain, suggesting 
that that the impact of three-dimensional dynamics is robust for astrophysical 
characterized by spatial and temporal scales that are much larger than kinetic scales.

In this article, we extend this study in several key ways. We begin by reviewing
the theory of particle acceleration in reconnection (section \ref{sec:theory}) and describing our kinetic
particle-in-cell (PIC) simulations (section \ref{sec:pic}). We then explore
the physics of two and three-dimensional reconnection to highlight the remarkable similarity of
many of the bulk properties (section \ref{sec:bulk}). In section \ref{sec:spectra}, we review
the physics of electron acceleration enhancement in 3D reconnection and demonstrate, by varying 
the spatial length in the third dimension, that transport and enhanced acceleration are intrinsically
linked. We show in section \ref{sec:guidefield} that a magnetic guide field plays an important role
in facilitating three-dimensional transport, and in section \ref{sec:conditions} introduce an `injection
criterion' that explains why energetic electrons are enhanced but protons are not. We discuss the astrophysical
implications of these results in section \ref{sec:discussion}.

\section{Particle Acceleration in the Guiding-Center Limit}
\label{sec:theory}

In order to examine electron acceleration
we assume a guiding-center approximation relevant for a 
strong guide field \citep{northrop63a,dahlin15a}. In this limit, the evolution of the kinetic energy $\epsilon$ 
of a single electron can be written as:
\begin{equation}
\label{eqn:particle}
\frac{d \epsilon}{d t} = q E_\parallel v_\parallel
  + \frac{\mu}{\gamma}\left( \frac{\partial B}{\partial t} + \ue \boldsymbol{\cdot} \boldsymbol{\nabla} B \right)
  + \gamma m_e v_\parallel^2 (\ue \boldsymbol{\cdot} \boldsymbol{\kappa})
\end{equation}
where $E_\parallel = \mathbf{E} \boldsymbol{\cdot} \mathbf{b}$ is the parallel electric field,
$\mu = m_e \gamma^2 v_\perp^2/2B$ is the magnetic moment, $\ue = c\mathbf{E} \cross \mathbf{B}/B^2$,
and $\boldsymbol{\kappa} = \mathbf{b} \boldsymbol{\cdot} \nabla \mathbf{b}$ is the magnetic curvature.
The velocity components parallel and perpendicular to the magnetic field are
$v_\parallel$ and $v_\perp$, respectively; $\gamma$ is the relativistic Lorentz factor, and
$\mathbf{b}$ is the unit vector in the direction of the local magnetic field.

The first term on the right-hand-side of the equation corresponds to acceleration
by parallel electric fields, which are typically localized near the reconnection 
X-line and along separatrices. The second term corresponds to betatron acceleration
associated with $\mu$ conservation in a temporally and spatially varying magnetic field.
Because reconnection releases a system's magnetic energy, this typically causes electron
cooling \cite{dahlin14a}.
The last term corresponds to Fermi reflection of particles from contracting magnetic field lines
\citep{drake06a,drake10a,hoshino12a,dahlin14a}. 
Both $E_\parallel$ and Fermi reflection change the parallel energy 
of the particles, while betatron acceleration changes the perpendicular energy.
The term $\uek$ corresponds to local field line contraction: $\uek = - \dot{\ell}/\ell$ (where $\ell$ is
the field line length) and is linked to the conservation 
of the parallel adiabatic invariant $\int v_\parallel d\ell$.
The guiding-center approximation given in Eq.~(\ref{eqn:particle}) is accurate
when electrons are well-magnetized. In the weak-guide field regime, other terms
such as the polarization drift may be significant (compare Li et al.\cite{li15a}).
However, the polarization drift gives the change in the electron bulk flow
energy which is typically small for a realistic electron-to-ion mass ratio.

\section{Particle-in-Cell Simulations}
\label{sec:pic}

We explore particle acceleration in reconnection via simulations using the massively parallel 3D
particle-in-cell (PIC) code {\tt p3d} \citep{zeiler02a}. Particle trajectories
are calculated using the relativistic Newton-Lorentz equation, and the electromagnetic
fields are advanced using Maxwell's equations. The time and space coordinates are 
normalized, respectively, to the proton cyclotron time $\Omega_{ci}^{-1} = m_i c/eB_{x0}$ 
and inertial length $d_i = c/\omega_{pi}$. 
The typical grid cell width $\Delta = d_e/4$, where $d_e = d_i \sqrt{m_e/m_i}$ is the electron
inertial length. 
The time step is $dt = \Omega_{ce}^{-1}/4$, where
$\Omega_{ce} = (m_i/m_e)\Omega_{ci}$ is the electron cyclotron frequency.

All simulations are initialized with a force-free configuration and use
periodic boundary conditions. This is chosen as the most generic model
for large-scale systems such as the solar corona where the density jump
between the current layer and upstream plasma is not expected to be important.
The magnetic field is given by $B_x =
B_{x0} \tanh(y/w_0)$ and $B_z = \sqrt{(1+b_g^2)B_{x0}^2-B_x^2}$,
corresponding to an asymptotic guide field $B_{z0}= b_g B_{x0}$.
We include two current sheets at $y=L_y/4$ and $3L_y/4$ to produce a periodic
system, and choose $w_0 = 1.25d_e$.
This initial configuration is not a kinetic equilibrium, which
would require a temperature anisotropy \cite{bobrova01a}, but is in pressure
balance.
We use at least 50 particles per cell per species. 
The initial electron and proton temperatures are equal and isotropic with
$T_0 = 0.25m_i c_A^2$, and the initial density
$n_0$ and pressure $P$ are constant so that $\beta_x = 8\pi P/B_{x0}^2
= 0.5$. The speed of light is $c = 3 c_A \sqrt{m_i/m_e}$, where
$c_{A}=B_{x0}/\sqrt{4\pi m_i n_0}$ is the Alfv\'en speed based on the
reconnecting component of the magnetic field.


Table \ref{tab:simulations} lists the simulation configurations 
discussed in this paper. We focus on configurations SM (`medium') and SL (`large') with
spatial dimensions $L_x \times L_y = 51.2 d_i \times 25.6 d_i$ and $L_x \times L_y = 102.4 d_i \times 51.2 d_i$, respectively. 
The larger simulations (SL) have more magnetic flux to reconnect, and can therefore run for a longer
time and generate many more energetic electrons. 
However, the SM configuration is $\sim8$ times less expensive and therefore better suited 
for parameter scans. Except where noted, the simulations are performed with a guide field $b_g = 1$ and the three-dimensional simulations use 
$L_z/d_i = 25.6$ (as in Dahlin et al., 2015\cite{dahlin15a}). The simulation with $b_g = 1.5$ in configuration SM
uses $\Delta/d_e = \Omega_{ce} dt = 1/6$. The electron-positron (pair plasma) configuration (S1) uses $\Omega_{ce} dt = 1/20$ and $\Delta/d_e = 1/4$.

The PIC formulation results in some numerical heating that can differ between 2D and 3D domains.
To simplify comparisons of particle energization, we perform a set of simulations with $L_z \sim 1.6 d_i \ll L_x, L_y$
so that the numerical heating is the same as in 3D, yet the reconnection physics remains essentially 2D 
(see Section \ref{sec:spectra} and Fig.~\ref{fig:lz}).
To reduce computational expense, configurations SM and SL have an artificial proton-to-electron mass ratio $m_i/m_e = 25$.
Simulations with $m_i/m_e = 100$ (S100) and $m_i/m_e = 1$ (S1) are also presented to explore the impact of the separation between
electron and proton scales.

\begin{table}
\begin{tabular}{| l || c | c | c | c | c |}
Name & $L_x/d_i$ & $L_y/d_i$ & $L_z/d_i$ & $m_i/m_e$ & $b_g$ \\
SM   &  51.2 & 25.6 & 2D, 1.6 - 25.6  & 25  & 0 - 1.5 \\ 
SL   & 102.4 & 51.2 & 2D, 1.6, 25.6 & 25  & 0, 0.5, 1.0 \\
S1   & 102.4 & 51.2 & 1.6, 51.2     & 1   & 1.0\\
S100 &  51.2 & 25.6 & 0.8, 25.6     & 100 & 1.0 \\
\end{tabular}
\caption{Simulation Parameters.}
\label{tab:simulations}
\end{table}

\section{Overview of reconnection in 2D and 3D domains}
\label{sec:bulk}

Reconnection develops from particle noise via the tearing instability, which
generates interacting flux ropes that grow and merge until they reach the
system size. These tearing modes grow wherever 
$\mathbf{k} \boldsymbol{\cdot} \mathbf{B} = 0$ (Ref.~\onlinecite{furth63a}). In a
slab equilibrium with $\mathbf{B} = B_x(y) \mathbf{x} + B_z(y) \mathbf{z}$, such modes are
characterized by a wavevector $\mathbf{k} = k_x\mathbf{x} + k_z\mathbf{z}$ and grow on flux 
surfaces defined by $B_x/B_z = -k_z/k_x$. The pitch of the unstable mode is given by
a characteristic angle  $\theta = \arctan(k_z/k_x)$ with respect to
the reconnecting field $B_x$. In a two-dimensional system where $k_z = d/dz = 0$, tearing modes 
are constrained to grow at the center of the current sheet where $B_x = 0$.
However, in three-dimensional systems the nonlinear interaction of modes with different
pitches destroys flux surfaces and generates a stochastic, turbulent magnetic 
field that facilitates enhanced particle transport\cite{rechester78a,daughton11a}.

\begin{figure}
 \includegraphics{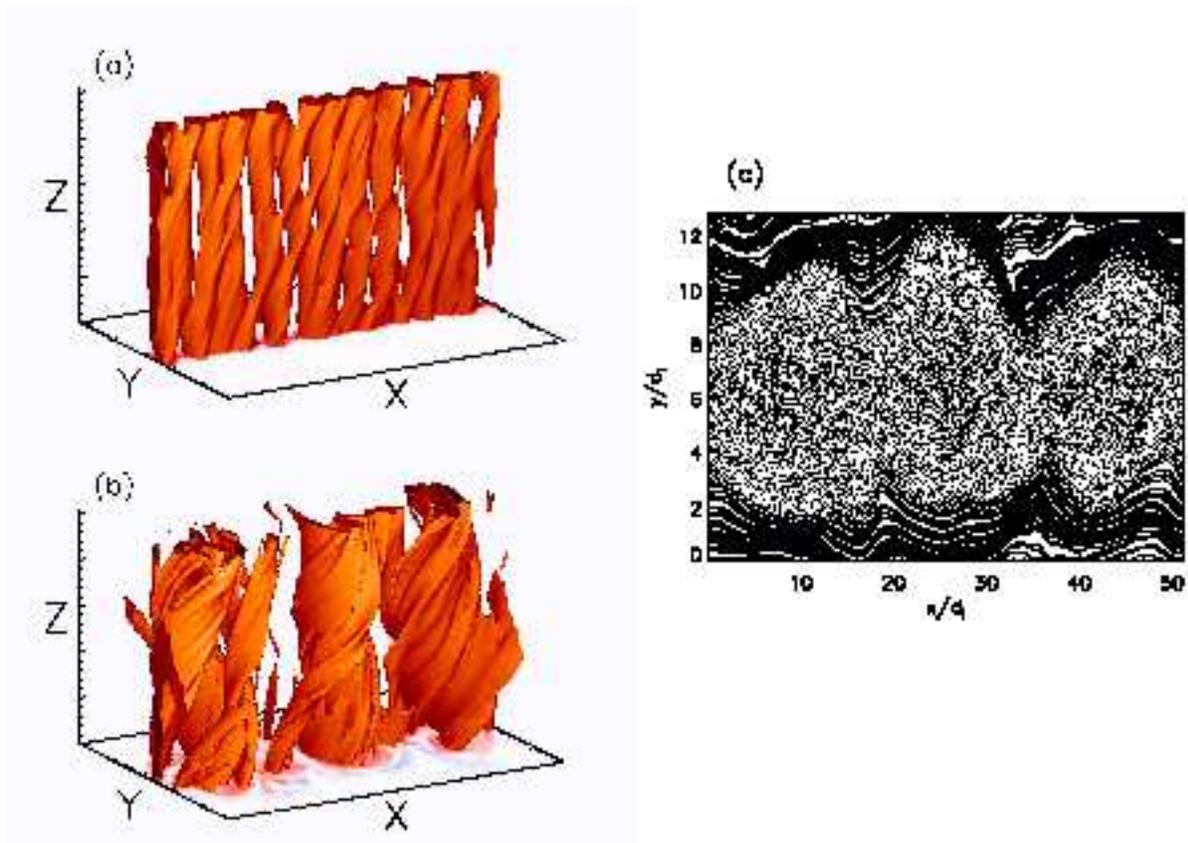}
 \caption{Results for configuration SM. (a,b) Isosurfaces of $J_{ez}$ at $\Omega_{ci}t = 12, 50$ respectively, 
illustrating the nonlinear filamentary current structure. 
(c) Poincar\'{e} surface-of-section at $\Omega_{ci} t = 50$. 
We trace a set of field lines beginning at $x = 0$, $0 < y < 12.8$, $z = 0$ and plot where 
they puncture the plane $z = 0$. The surface-of-section shows a clear boundary between the 
stochastic field lines inside the reconnecting region and the laminar unreconnected fields.
\label{fig:isojez_all}}
 \end{figure}

This stochastic magnetic structure is illustrated in Figs. \ref{fig:isojez_all}a-b, which show isosurfaces
of one component of the electron current density $J_{ez}$ in configuration SM.
At $t=12$, several tearing 
structures with $k_z \neq 0$ are visible (compare Fig.~1 in Daughton et al. 2011\cite{daughton11a}). 
The filamentary current distribution at $t=50$ showcases the late-time nonlinear development.
A different view of the filamentary structure, which emphasizes the stochastic
structure of the magnetic field, can be observed in the Poincar\'{e} surface-of-section 
shown in Figure \ref{fig:isojez_all}c.
There is a clear boundary
between the stochastic reconnecting region (disordered punctures) and the asymptotic,
laminar field. 

\begin{figure}
 \includegraphics[width=6in]{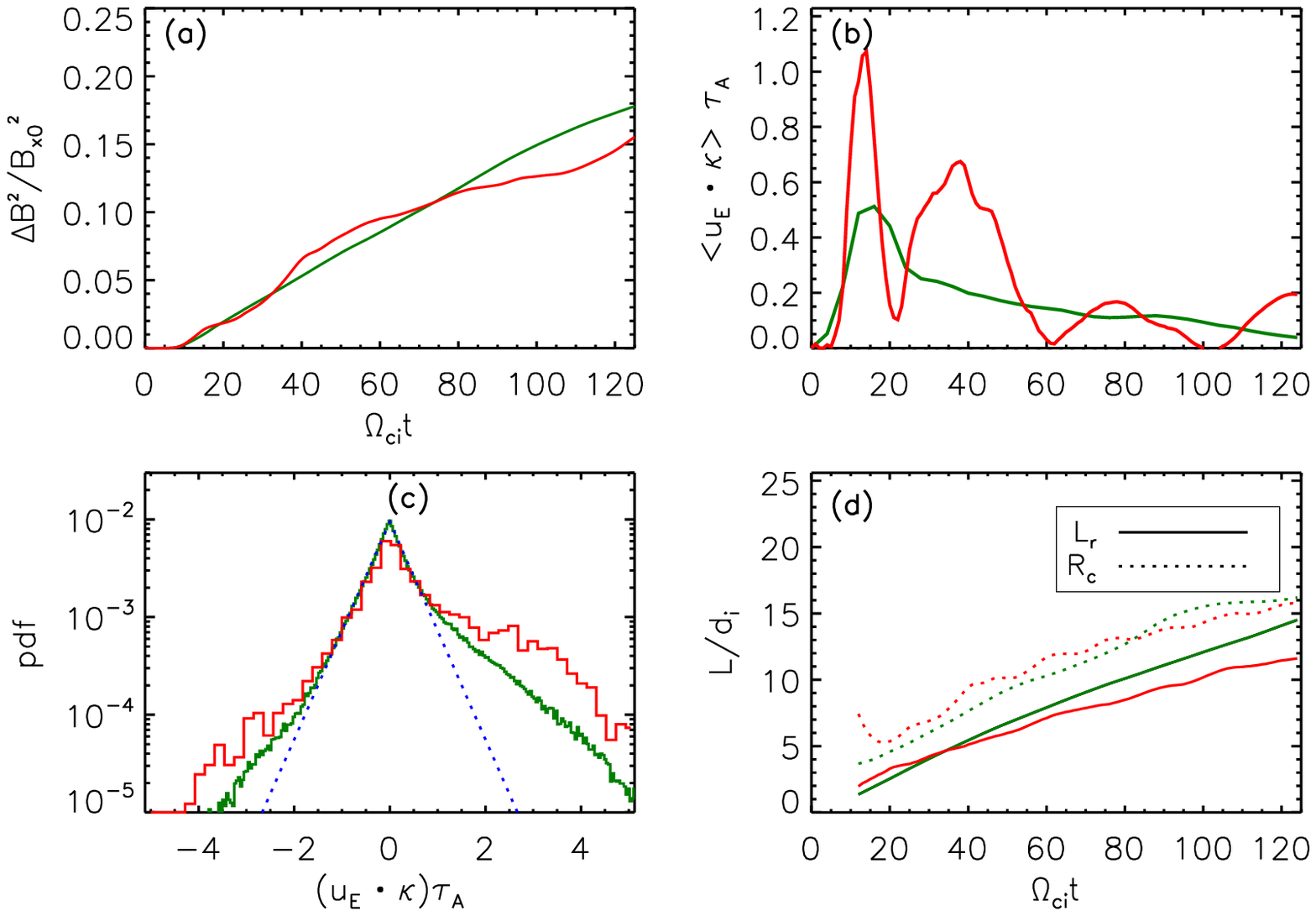}
 \caption[Magnetic Energy vs. Time in 2D and 3D simulations.]
{Results for configuration SL: 3D (green) and 2D (red).
(a) Magnetic energy release vs. time. 
(b) Spatially averaged $(\uek) \tau_A$ over the reconnecting region, where $\tau_A = 
L_x/c_A$ is the Alfv\'en crossing time.
(c) Probability distribution function of $(\uek) \tau_A$ at $\Omega_{ci}t = 40$, 
with a superimposed double-exponential fit (blue dotted line).
(d) Reconnection region half-width $L_r$ and radius of curvature $R_c$.
\label{fig:bulk}}
\end{figure}

The stochastic 3D dynamics do not substantially impact the magnetic energy release 
(see Fig.~\ref{fig:bulk}a), as has been noted by Daughton et al.\cite{daughton14a}
Another diagnostic for the energy release is the field-line contraction
$\uek$ that drives Fermi acceleration according to Eq.~(\ref{eqn:particle}). The spatial average $\ave{\uek}$, (
calculated over the stochastic reconnection region described below) is shown in Fig.~\ref{fig:bulk}b.
Although the 2D simulation is relatively bursty, the overall time-evolution is
comparable to that in 3D. 
In both simulations, $\left<\uek \right>$ decreases in time as islands grow and the
typical radius of curvature $R_c = |\boldsymbol{\kappa}|^{-1}$ increases.
Figure \ref{fig:bulk}c shows the probability distribution function (pdf)
of $\uek$ inside the stochastic reconnecting region at $\Omega_{ci}t = 40$. 
The pdf in the 3D system is symmetric for small values of $\uek$, and 
is consistent with a double-exponential distribution $\propto \exp[-|\uek|\tau_A/0.39]$ where
$\tau_A = L_x/c_A$ is the Alfv\'en crossing time.
The symmetric component can only produce net acceleration through a second-order Fermi process.
The positive mean value
$\ave{\uek} \tau_A \approx 0.2$ is due to the large positive tail visible for $(\uek) \tau_A > 1$.
Although the characteristic scales for first and second-order
components are comparable, the first-order mechanism is far more efficient, 
and hence is the dominant driver of particle acceleration in this system. 
The symmetric (second-order Fermi) component is consistent with Alfv\'enic fluctuations 
where the flow and curvature are out of phase corresponding to no net field-line contraction, 
i.e. $\ave{\uek} = 0$.

\begin{figure}
 \includegraphics[width=6in]{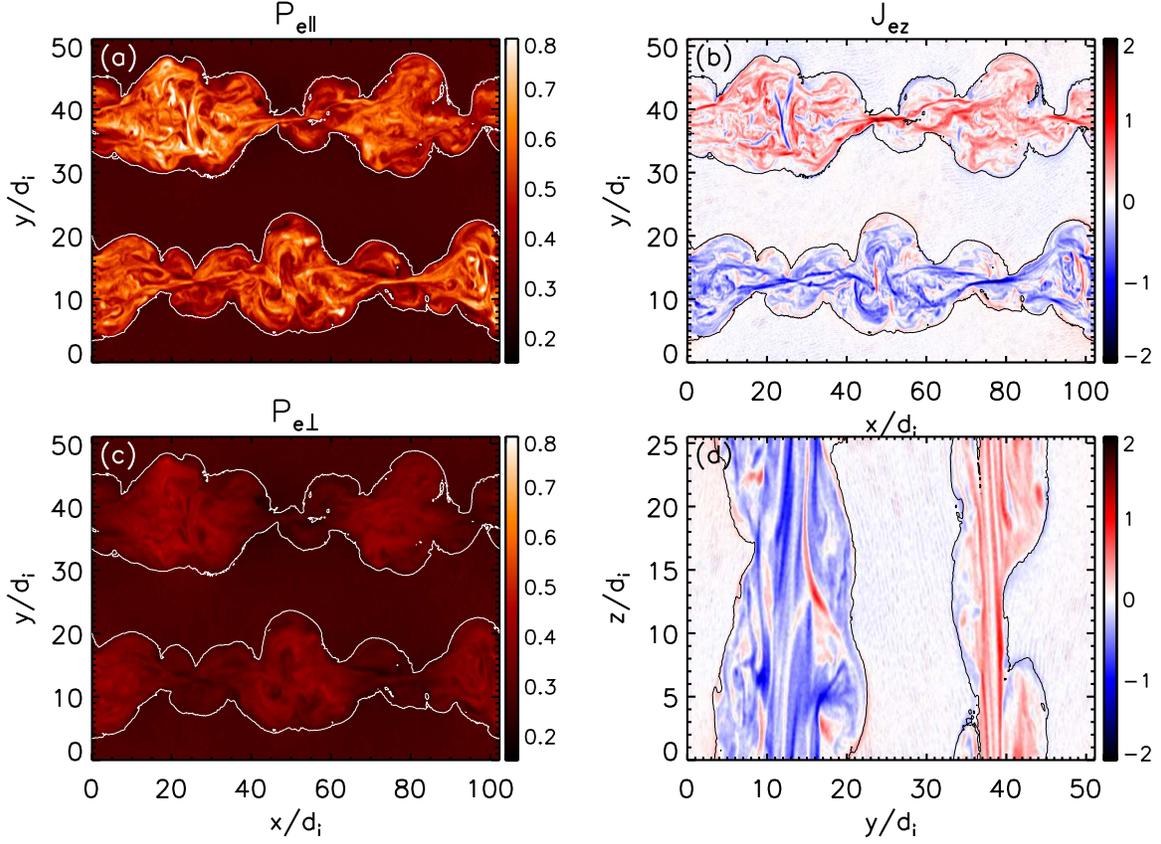}
 \caption
{Results for configuration SL.
(a)(c) $z=0$ slices at $\Omega_{ci}t = 100$ of
parallel and perpendicular electron pressure. 
(b) (d) $z$-directed electron current density in planes defined by
$z=0$ and $x=0$, respectively. White and black outlines 
indicate the contour $P_{e\parallel,nt} = 0.04n_0 T_{e0}$, a marker
for the region of reconnected magnetic field.
\label{fig:mixing}}
\end{figure}

\begin{figure}
 \includegraphics[width=6in]{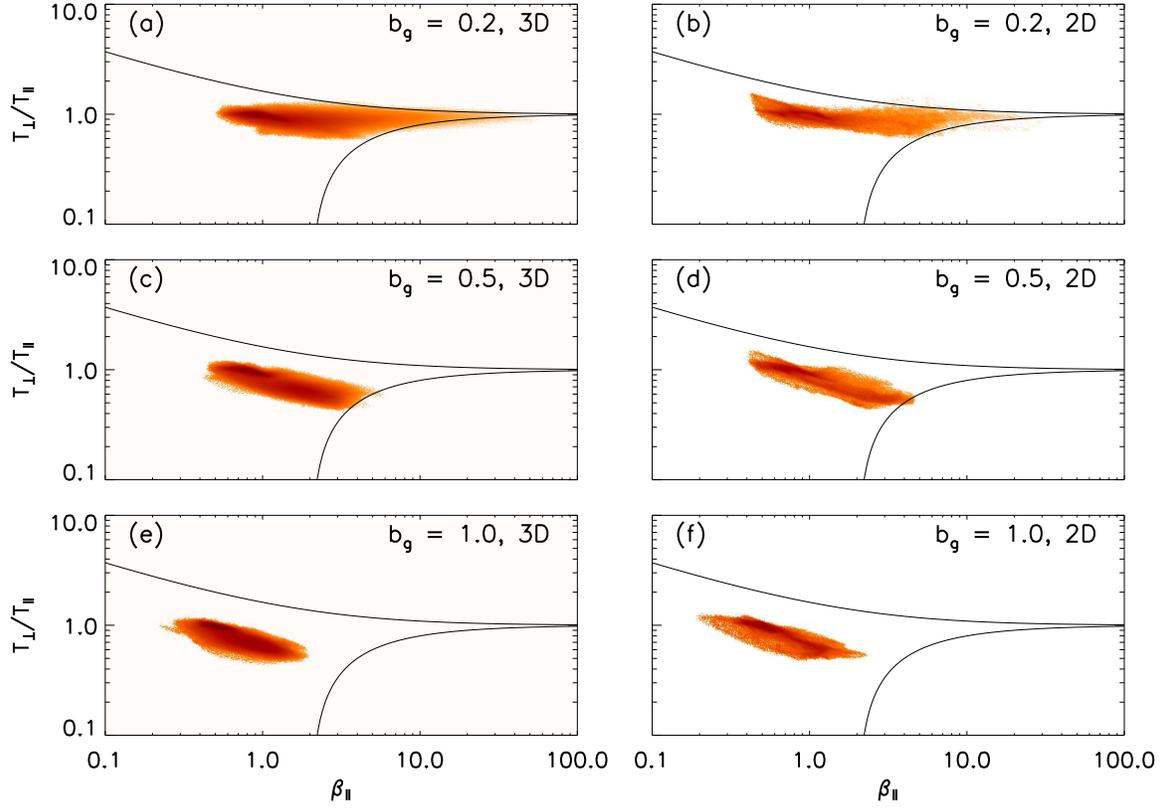}
 \caption
{Results for configuration SM. Phase space distribution of 
temperature ratio $T_\perp/T_\parallel$ vs. $\beta_\parallel$. Black lines
indicate marginal stability conditions for the ideal firehose (bottom) 
and mirror (top) instabilities.
\label{fig:firehose}}
\end{figure}

It has been shown previously that the development of pressure anisotropy with 
$P_\parallel \gg P_\perp$ causes the cores of magnetic islands to approach firehose 
marginal stability, where the tension driving magnetic reconnection ceases, thereby 
throttling reconnection \cite{drake06a,drake10a,schoeffler11a}. 
Figure \ref{fig:mixing}a,c show that a significant anisotropy $P_\parallel > P_\perp$
develops, as is the typical case in 2D (not shown). This suggests that the plasma heating and energization
occurs in similar ways in 2D and 3D, and that the turbulent magnetic field generated in 3D does 
not isotropize the plasma. Phase space plots of temperature anisotropy $T_\parallel/T_\perp$ and $\beta_\parallel = 8 \pi P_\parallel/B^2$
are shown in Fig.~\ref{fig:firehose} for three values of the guide field $b_g$ in 2D and 3D, along with 
marginal stability boundaries for the firehose and mirror instabilities (bottom and top, respectively)
in a similar format to that used previously in analyzing solar wind data \cite{kasper02a,bale09a}. 
For the strong guide field cases (\ref{fig:firehose}e,f) 
there is insufficient free energy in the reconnecting field to drive the system either firehose or
mirror unstable. However, for the simulations with $b_g = 0.2, 0.5$ (Fig.~\ref{fig:firehose}a-d)
the phase space brushes up against the stability boundaries. Hence, feedback of heating on reconnection 
via anisotropy-driven instabilities can occur in 3D systems.

Reconnection primarily drives electron acceleration parallel to the local magnetic field, generating
superthermal particles that fill the stochastic reconnecting region\cite{dahlin15a}. The
presence of reconnection-accelerated electrons is therefore a useful proxy for
the reconnection region. This is similar to the ``electron mixing" described by
Daughton et al. (Ref. \onlinecite{daughton14a}). Indeed, the volume defined by $P_{e\parallel,nt} \geq 0.04 n_0T_{e0}$ (where
$P_{e\parallel,nt}$ is the parallel energy density of electrons exceeding $\epsilon = 0.2m_ec^2$)
corresponds well to the reconnecting region as indicated by the electron pressure and current density 
(see Fig.~\ref{fig:mixing}).
Using this marker for designating the reconnection domain allows us to estimate the reconnected volume $V_{r}$ 
and can be used to determine a characteristic width in the
3D system: $L_r = V_{r}/L_xL_z$. The 2D analogue is the area inside the separatrices of the primary
X-line.
A mean inflow velocity can then be determined from $v_{in} = \dot{L_r}/2$, yielding comparable
$v_{in}/c_A \approx 0.057, 0.045$ for the 3D and 2D simulations respectively ($\dot{L_r}$ is averaged
from $\Omega_{ci}t=8$ to $\Omega_{ci}t=125$). A calculation of the 2D reconnection rate,
determined from the time rate of change of the flux function at the primary X-line, yields a nearly
identical inflow velocity $v_{in}/c_A \approx 0.049$.
Figure \ref{fig:bulk}d shows that the width of the reconnecting region is comparable to the mean radius of curvature,
consistent with $R_c \approx \ave{\kappa}^{-1} = B^2/|\mathbf{B} \cdot \grad \mathbf{B}| \approx (L_r/2) (B^2/B_x^2) 
\approx L_r$. In summary, the total magnetic energy conversion, field line contraction (Fermi drive), and reconnection
rate provide strong evidence for the remarkable simlarity of bulk properties of the 2D and 3D reconnection.

\section{Electron Acceleration}
\label{sec:spectra}

\begin{figure}
 \includegraphics[width=6in]{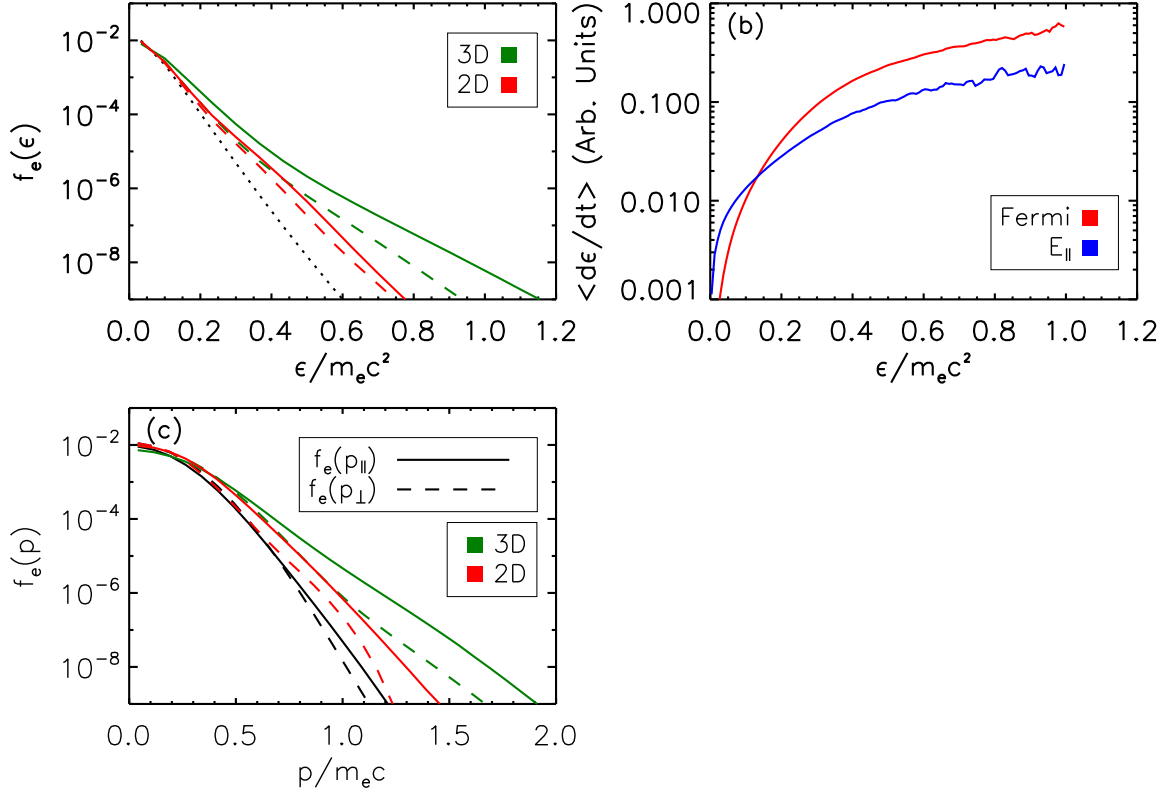}
 \caption[Magnetic Energy vs. Time in 2D and 3D simulations.]
{Results for configuration SL: 3D (green) and 2D (red).
(a) Energetic spectra at $\Omega_{ci} t = 50$ (dashed) and
$125$ (solid)
(b) Average energy gain due to Fermi acceleration (red) and $E_\parallel$ (blue) in
simulation 4a at $\Omega_{ci} t = 100$. 
(c) Parallel and perpendicular momentum
spectra at $\Omega_{ci}t = 0,125$.
\label{fig:all_accel}}
\end{figure}

In a previous study \cite{dahlin15a} we found that electron acceleration was enhanced 
in a 3D reconnecting system. These results are summarized in Fig.~\ref{fig:all_accel}.
Although there is substantial acceleration in both systems, the fraction
of electrons with energy exceeding $0.5 m_e c^2$ is more than an order of magnitude larger 
than in the 2D simulation (Fig.~\ref{fig:all_accel}a). 
However, as noted in Section \ref{sec:bulk}, the magnetic energy dissipation is comparable in 2D and 3D systems.
This suggests that the increased energetic electron production in the 3D system is due to enhanced
acceleration efficiency rather than an increase in the total energy imparted to
the plasma.
According to equation (\ref{eqn:particle}) the acceleration mechanisms have 
different scalings with the particle energy:
the Fermi reflection term is second-order in the parallel velocity, 
whereas the parallel electric field term is only first-order. 
The instantaneous average acceleration per particle 
for both $E_\parallel$ and Fermi
reflection in configuration SM, ($b_g = 1$) is shown in Fig.~\ref{fig:all_accel}b.
The bulk thermal electrons (low energies) 
are primarily accelerated by $E_\parallel$, whereas Fermi reflection
is more important at high energies. The energetic electrons are
primarily accelerated in the parallel direction so that the momentum
distribution $f(p_\parallel)$ exceeds $f(p_\perp)$ (Fig.~\ref{fig:all_accel}c), 
consistent with acceleration via Fermi reflection and $E_\parallel$. 
To summarize: electron acceleration, primarily driven by field-line contraction, 
is enhanced in 3D systems. However, this is not due to greater energy release, 
so must instead be due to enhancement of the acceleration efficiency.

\begin{figure}
 \includegraphics{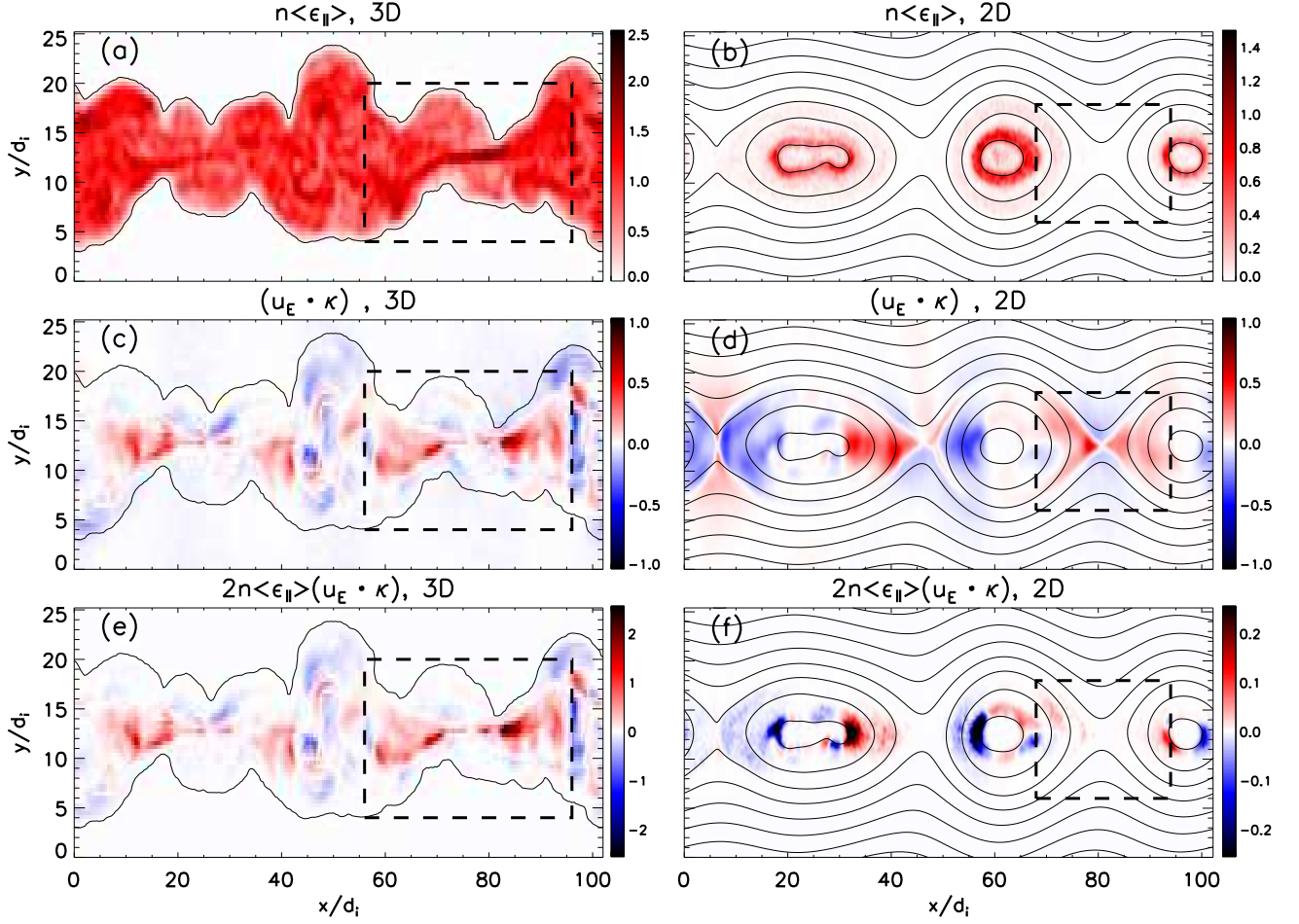}
 \caption{Results for configuration SL: 3D (left) and 2D (right) at $\Omega_{ci}t = 100$
in the plane $z=0$.
(a-b) Parallel energy density for electrons with $\epsilon > 0.5 m_ec^2$.
(c-d) Field-line contraction rate $\uek$.
(e-f) Fermi acceleration for electrons with $\epsilon > 0.5 m_ec^2$.
The similarity between (c) and (e) reflects the nearly uniform spatial
distribution of the energetic electrons in the reconnecting region.
Dashed boxes outline one X-line in each panel. 
\label{fig:heat3d_100}}
 \end{figure}

As was discussed in Section \ref{sec:bulk}, the stochastic structure of the 
magnetic field in 3D systems allows field-line-following particles to wander 
throughout the chaotic reconnecting region \cite{jokipii69a}. However, in 2D
systems reconnected field lines form closed loops (islands) that trap 
particles. The impact of topology on transport is reflected in the spatial 
distribution of the most energetic particles (shown in 
Fig.~\ref{fig:heat3d_100}a,b). These particles occupy narrow bands well inside 
the islands in the 2D simulation, but are distributed throughout the reconnecting 
region in the 3D simulation. 
The most efficient electron acceleration regions
are near the X-lines and at the ends of islands (Fig.~\ref{fig:heat3d_100}c,d).
In the 2D system, trapped energetic practicles are unable to access these regions,
and the overall acceleration efficiency is suppressed with respect to the 3D system where
the energetic particles wander the reconnecting region and undergo continuous acceleration.
Figure \ref{fig:heat3d_100}e,f shows the resulting rate of Fermi energization for electrons 
with $\epsilon > 0.5 m_e c^2$, revealing an order-of-magnitude difference between the two
systems.

\begin{figure}
 \includegraphics[width=6in]{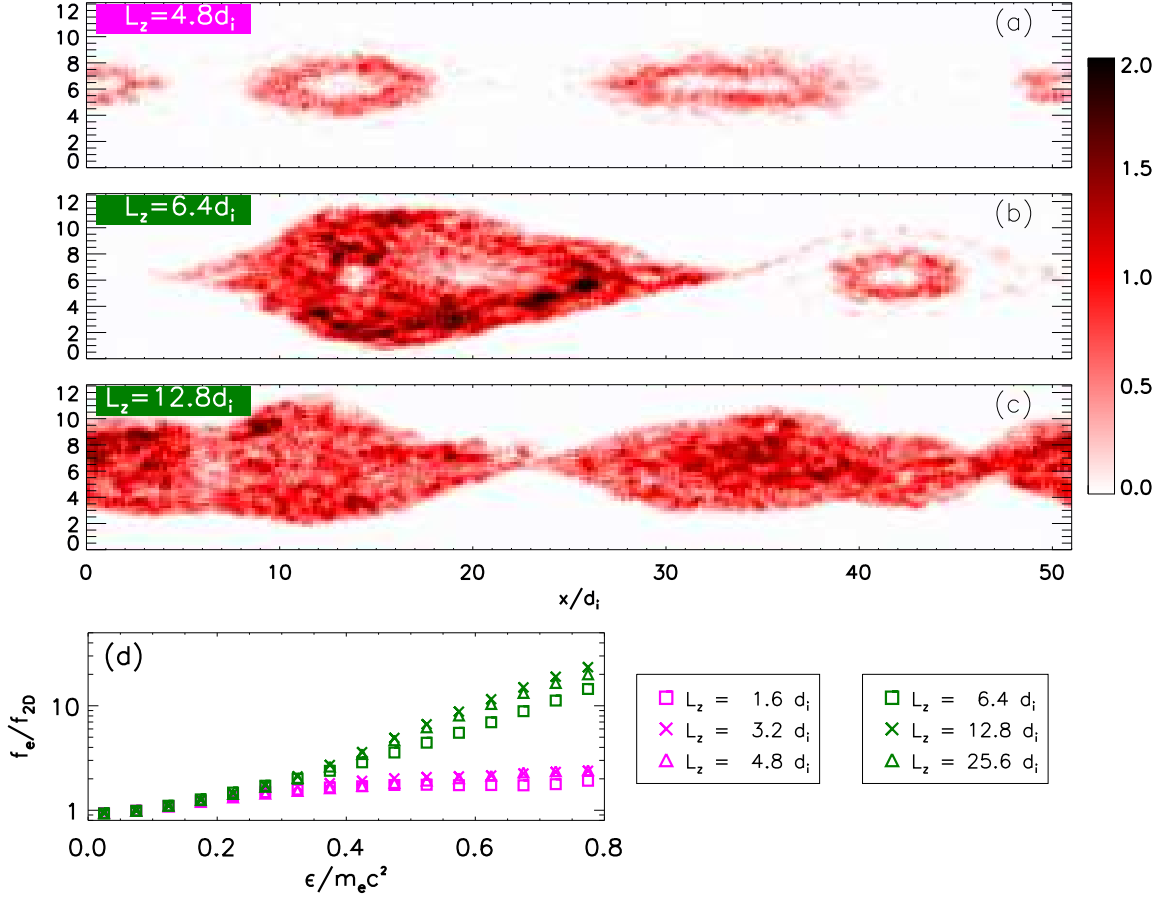}
 \caption
{Results for configuration SM, $b_g = 1$.
(a-c) Spatial distribution of electrons with
$\epsilon > 0.5 m_e c^2$.
(d) Electron energy spectra normalized to the 2D spectrum at $\Omega_{ci}t = 50$ .
\label{fig:lz}}
\end{figure}

In order to examine the transition from 2D to 3D reconnection, we performed a set of simulations
with different values of $L_z$,
from a quasi-2D system with $L_z = 1.6 d_i$ to a simulation with $L_z = L_y = 25.6d_i$. 
Figure \ref{fig:lz}a-c shows the spatial distribution 
of the energetic electrons ($\epsilon > 0.5 m_ec^2$) for several of these simulations.
Surprisingly, there is a sharp transition at $L_z = 6.4 d_i$: below this threshold, energetic electrons
are trapped inside islands, whereas above this threshold the energetic electrons are space-filling.
Electron energy spectra exhibit the same transition (Fig.~\ref{fig:lz}d).
Simulations with $L_z < 6.4 d_i$ do not show enhancement with respect to the 2D result, 
whereas simulations $L_z \geq 6.4 d_i$ are consistent with the 3D result. 
This reinforces the correlation between enhanced transport and acceleration in 3D systems.

\section{The Role of the Guide Field}
\label{sec:guidefield}
In a recent study of two-dimensional kinetic reconnection (Dahlin et al., 2016 \cite{dahlin16a}), we found 
that the magnetic guide field was a vital parameter that controls the 
efficiency of electron acceleration. In a system with a guide field much
smaller than the reconnecting component, the dominant electron 
accelerator was a Fermi-type mechanism that
preferentially energizes the most energetic particles. In the 
strong guide field regime, however, the field-line
contraction that drives Fermi reflection was negligible. Instead, 
parallel electric fields ($E_\parallel$) were primarily
responsible for driving electron heating. Electron acceleration 
was suppressed in the systems with a strong guide field. We argued 
that this was due to the the weaker energy scaling of $E_\parallel$ 
acceleration.

\begin{figure}
 \includegraphics[width=6in]{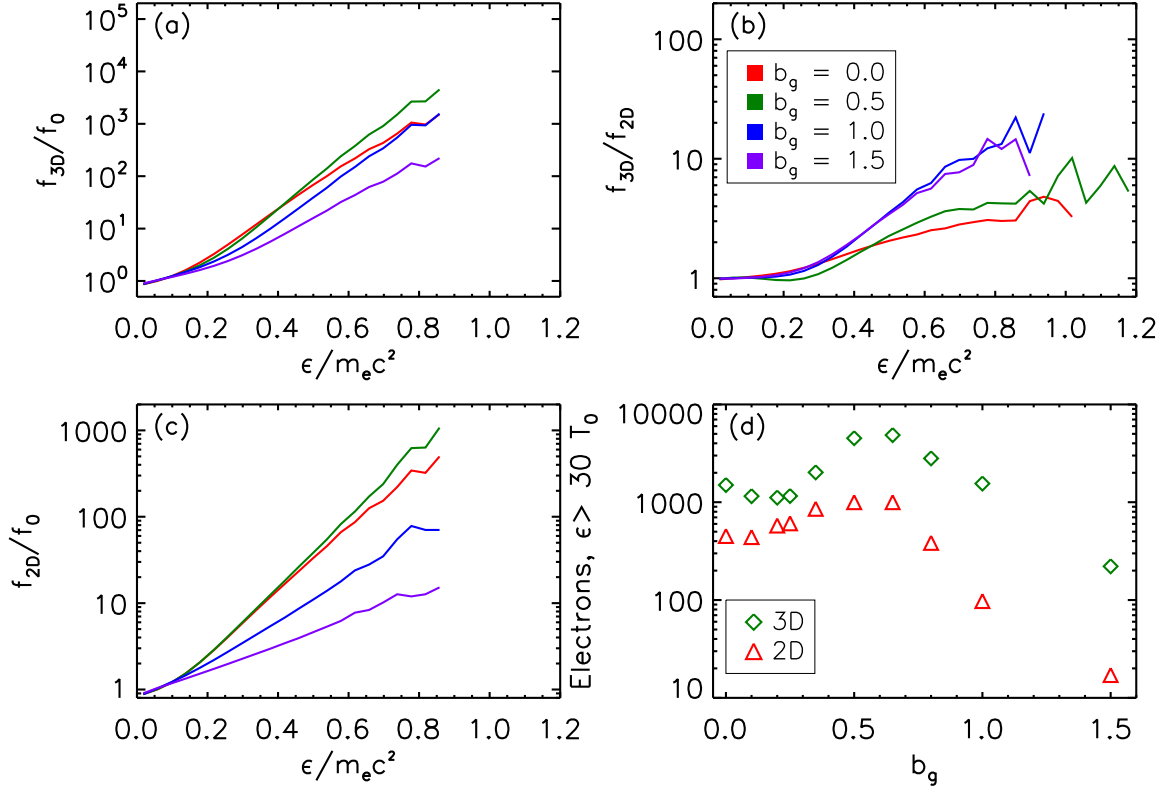}
 \caption
{Results for configuration SM at $\Omega_{ci}t = 50$.
Three-dimensional electron energy spectra $f_{3D}$ normalized to the initial spectrum $f_0$ (a) and quasi-2D spectra $f_{2D}$ (b).
(c) quasi-2D spectra $f_{2D}$ normalized to the initial spectrum $f_0$. (d) Electrons exceeding $30 T_{e0}$
vs. guide field. The system with $b_g = 0.65$ generates the greatest number of energetic electrons.
\label{fig:spectra_guide}}
\end{figure}

To probe the role of the guide field in three-dimensional transport and
particle acceleration, we performed several three-dimensional simulations 
in the configuration SM with $0 \leq b_g \leq 1.5$ and compared these
results with a set of quasi-2D simulations ($L_z = 1.6 d_i$).
Selected electron energy spectra from these simulations are shown in Figure \ref{fig:spectra_guide}a-c.
Figure \ref{fig:spectra_guide}d shows the number of electrons exceeding $30 T_{0}$, revealing that
energetic electron production varies strongly with the guide field.
The efficiency of the Fermi mechanism that drives energetic electron production weakens with increasing guide field\cite{dahlin16a}. 
The quasi-2D spectra (Fig.~\ref{fig:spectra_guide}c) are consistent with this result, explaining why
the energetic electron production diminishes as $b_g > 1$. The decreasing energetic electron production
for $b_g \ll 1$ must then be due to three-dimensional effects. Indeed, Fig.~\ref{fig:spectra_guide}b shows that the relative 
3D enhancement ($f_{3D}/f_{2D}$) increases with $b_g$ until it saturates above $b_g = 1$ (there is little difference
between $b_g = 1, 1.5$).


Representative field lines for several of the 3D simulations are shown in Fig.~\ref{fig:tubes}, 
illustrating the differing field line structures (Fig.~\ref{fig:tubes}a shows field lines from
a 2D simulation).
In Fig.~\ref{fig:tubes}b, the field lines are clearly stochastic, and do not show clear
flux rope structures that would trap particles; hence, electrons are free
to return to acceleration regions.
In weak guide-field reconnection, however (Figs. \ref{fig:tubes}c-d), field lines wrap 
around clearly-defined flux ropes, and in the case with $b_g = 0$, the field lines approximately
close on themselves. This structure inhibits particle escape from islands, similar to the 2D
structure shown in Fig.~\ref{fig:tubes}a, where flux surfaces are closed and particles become
topologically disconnected from acceleration regions. In 3D, the guide field plays a role in
breaking the 2D symmetry and allowing particles to escape along the flux rope axis. This
explains why the three-dimensional enhancement increases with the guide field.
The saturation above $b_g = 1$ can be explained by noting that magnetic structures are typically 
elongated along the guide field for $b_g > 1$ so that particles must move farther along the axis to escape
the island.
Hence, further 3D enhancement over what is shown in Fig.~\ref{fig:spectra_guide}b should not occur.

\begin{figure}
 \includegraphics[width=6in]{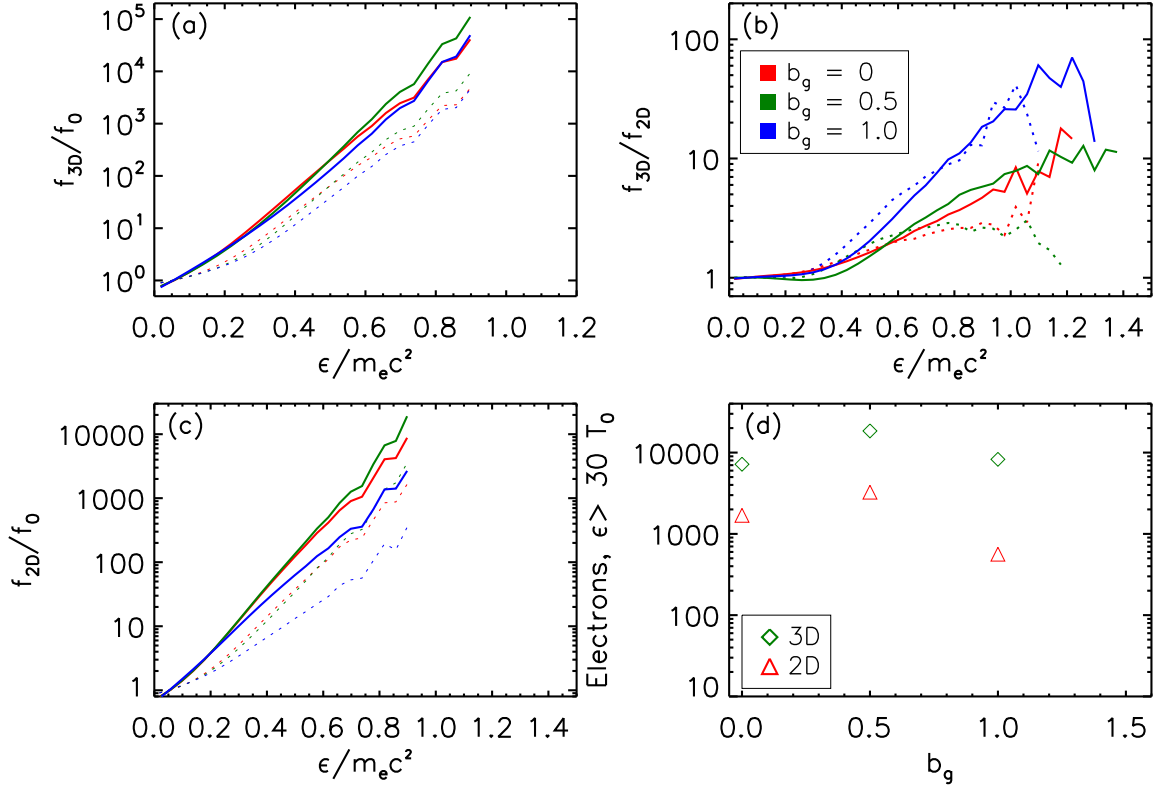}
 \caption
{Results for configuration SL.
Three-dimensional electron energy spectra normalized to the initial spectrum (a) and quasi-2D spectra (b)
at $\Omega_{ci}t = 125$ (solid) and $\Omega_{ci}t = 50$ (dashed). 
(c) quasi-2D spectra normalized to the initial ($t = 0$) spectrum. (d) Electrons exceeding $30 T_{e0}$ at $\Omega_{ci}t = 50$
vs. guide field. The system with $b_g = 0.5$ generates the greatest number of energetic electrons.
\label{fig:spectra_guide_l}}
\end{figure}

\begin{figure}
 \includegraphics{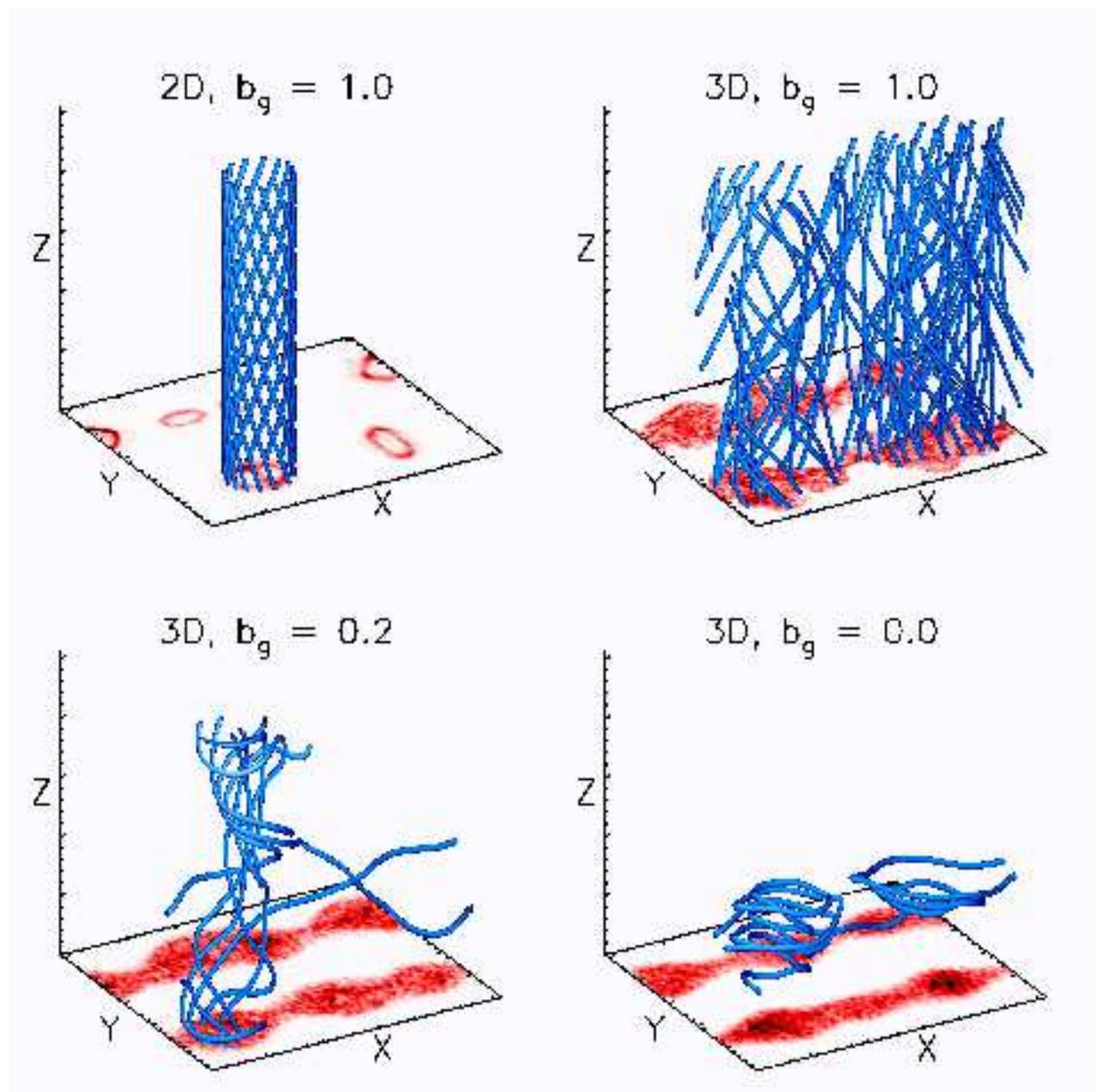}
 \caption{Results from configuration SM.
Single field lines (blue) for simulations with different guide fields.
Two-dimensional ($z = 0$) slices of the energetic electron density are shown at the bottom of each panel.
\label{fig:tubes}}
 \end{figure}

The convolution of the Fermi acceleration efficiency and the effectiveness of three-dimensional transport
results in a peak energetic electron production at $b_g \approx 0.6$. Results from a  set of 3D simulations with
$L_x \times L_y \times L_z = 102.4 \times 51.2 \times 25.6$ (configuration SL) are shown in Fig.~\ref{fig:spectra_guide_l} (dashed
lines indicate an earlier time). These simulations show that the enhancement $f_{3D}/f_{2D}$ increases as the
spectra extend to higher energies, suggesting that three-dimensional transport will be even more important in
physical systems such as the corona where the length scales $L \gg d_i$. The most efficient guide field, in
these simulations $b_g \approx 0.6$, will likely depend both on the system size and on other plasma parameters 
such as the plasma beta, which can impact the relative efficiency of Fermi and $E_\parallel$-driven acceleration.

\section{An `Injection Criterion' for Enhanced Acceleration}
\label{sec:conditions}
\begin{figure}
 \includegraphics{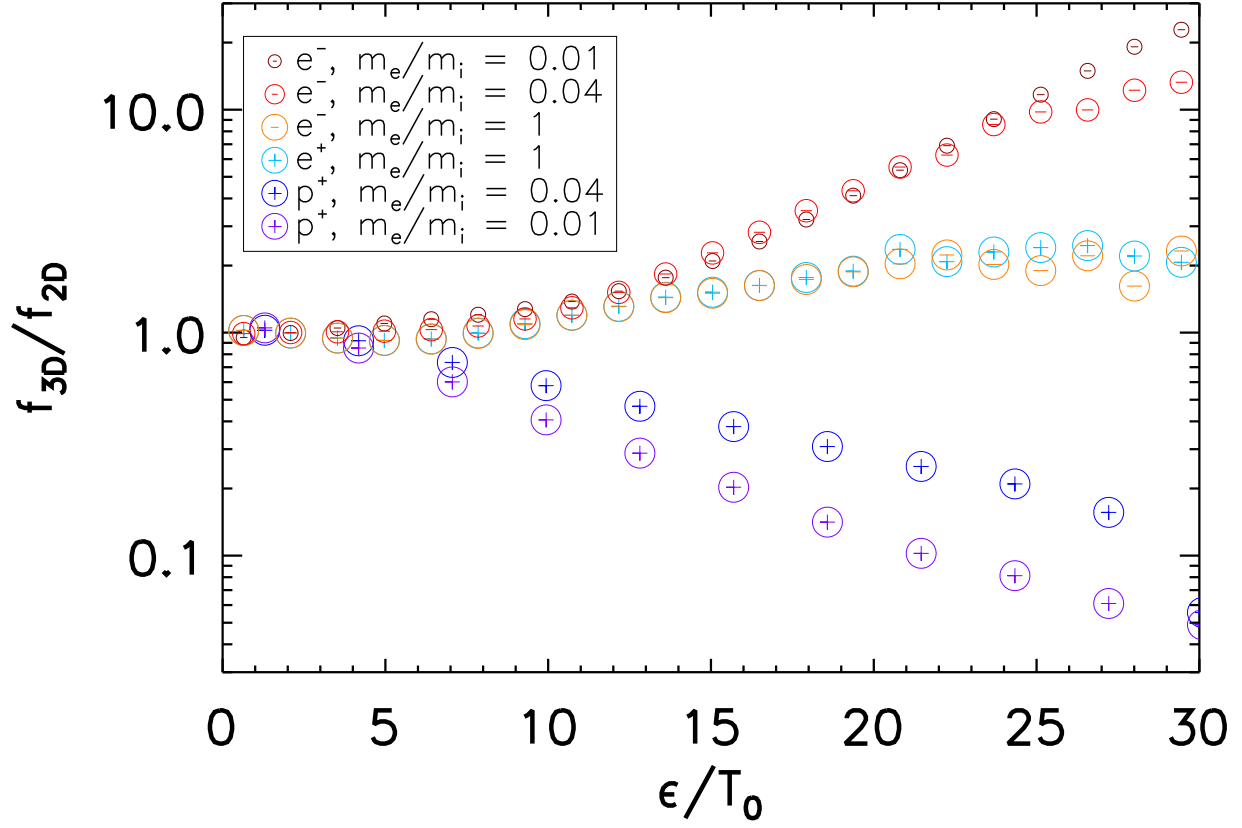}
 \caption{
Results from configurations SM, S1, and S100.
Three-dimensional enhancement $(f_{3D}/f_{2D}$) for simulations with different mass ratios $m_i/m_e$.
For the pair plasma ($m_i/m_e =1$), both species exhibit a small enhancement $\sim 2$,
whereas for simulations with $m_i \gg m_e$, energetic electrons are enhanced whereas energetic ions
are suppressed.
\label{fig:spectra_inj}}
 \end{figure}

A limitation of the present simulations is the use of an artificial mass ratio, which reduces the separation 
between proton and electron scales. To examine how the mass ratio impacts particle acceleration, 
we performed simulations with $m_i/m_e = 1, 25, 100$ (configurations S1, SM, and S100) and $b_g = 1$.
Figure \ref{fig:spectra_inj} shows the relative enhancement of the energy spectra in the three-dimensional
simulations ($f_{3D}/f_{2D}$). 
For the electron-positron case ($m_i/m_e = 1$), there is only a slight enhancement ($\sim 2$) in the energetic
tail for both species. For the electron-proton cases ($m_i/m_e = 25, 100$), the energetic electrons are enhanced
whereas the energetic ions are suppressed. 
The enhancement (suppression) of the energetic electrons (ions) is greater for the more realistic
mass ratio. This trend, along with the weak enhancement for the electron-positron case, suggesting that the 
separation of scales between species plays an important role in 3D particle acceleration and that the
impact of 3D transport should be robust for the physical mass ratio.

We propose the following explanation: in order for a charged particle to accelerate multiple times,
it must propagate upstream against the Alfv\'enic outflow that ejects plasma from the energy release
regions near the X-line and at the ends of islands. The condition $v/c_A \gg 1$ then acts as an 
`injection criterion' for efficient acceleration (this is analogous to the injection problem in
shock-driven particle acceleration). Heavy species (protons for $m_i/m_e > 1$, and both 
electrons and positrons for $m_i/m_e = 1$) are responsible for the bulk inertia and hence the 
reconnection outflow velocity $c_A$. The characteristic velocity of these particles is therefore
of the same order as the Alfv\'en speed ($v \sim c_A$), so that the bulk particles do not meet 
the injection criterion and do not experience enhanced 3D acceleration.
A few particles in the tail of the distribution satisfy the criterion in the electron-positron case,
explaining the small ($\sim 2$) enhancement at high energies.
For the electron-proton simulations ($m_i/m_e = 25, 100$) the electron thermal velocity $v_{th,e}/c_A \sim \sqrt{\beta_x (m_i/m_e)} \gg 1$ greatly
exceeds the Alfv\'en speed.
Recent numerical studies suggest that even in environments with $\beta \ll 1$, reconnection heats 
electrons to an appreciable fraction of the available magnetic energy density \cite{shay14a,haggerty15a}:
$\Delta T_e\approx 0.03 m_e c_{Ae}^2$, corresponding to $v_{th,e}/c_A \approx \sqrt{0.06m_i/m_e} \approx 10$
so that essentially all reconnection-heated electrons will satisfy the criterion, independent of the initial 
temperature. In contrast, ions are typically sub-Alf\'enic and would require an injection mechanism (e.g. \cite{drake09a,drake09b,knizhnik11a})
to undergo continuous acceleration. However, the suppression of ion acceleration in 3D systems
is surprising. While the relative increase in energy going into energetic electrons may play some role, it is
not clear that this should preferentially impact the energetic ions. Further treatment is 
beyond the scope of this paper.

The injection criterion may require modification in the large guide field limit ($B_z/B_0 \gg 1$).
In this limit, the electrons dominantly stream in the $z$-direction, so the relevant velocity
to compare to the outflow speed is is $v_x \approx v B_x/B \approx v/b_g$ so the injection criterion
becomes $v/c_A \gg b_g$. In strongly relativistic systems, 
all velocities approach $c$, so that the injection criterion cannot be met. 
This suggests that enhanced 3D acceleration should not occur for relativistic reconnection in 
either pair \cite{guo14a,sironi14a} or electron-proton plasmas.


\section{Discussion}
\label{sec:discussion}

Electron acceleration in three-dimensional systems is a complex problem that intrinsically
depends on the transport properties of reconnection-generated magnetic fields.
The picture that emerges from this set of simulations is that
particle acceleration is efficient in a three-dimensional system when the energetic
population can freely access acceleration sites and thereby achieve continuous energy gain. 
This requires both topological access to energy release regions and a super-Alfv\'enic 
particle velocity in order to explore the open topology at a faster time scale than the system evolves 
(most easily understood as the ejection of flux from the X-line). 
While electrons satisfy this condition rather easily, heavy species such as protons would
require an injection mechanism in order to be able to propagate upstream
against the reconnection outflow.

Efficient transport requires a strong guide field. The field structure in antiparallel
reconnection is quasi-laminar, so that energetic particles are still well-trapped in islands.
Propagation upstream against the Alfv\'en velocity is not possible in strongly relativistic 
reconnection, where all characteristic velocities approach the speed of light.
This is consistent with studies by Guo et al.\cite{guo14a,guo15a,guo16a},
that exhibit no substantial difference in energetic particle production 
between two and three-dimensional simulations in the relativistic regime. 

The nonthermal electron spectra in both simulations do not assume a power law form
as is frequently observed in nature. The maximum energy gain is limited due to the
modest size of the simulations; previous 2D simulations have shown that
the total energy gain is greater in larger systems \cite{dahlin14a}.
An additional issue is that these simulations have periodic boundary conditions that
prevent particle loss from the system. 
Solar observations suggest that electrons are confined in regions of energy release 
in the corona \cite{krucker10a}; possible mechanisms for this confinement include
mirroring and double layers\cite{li12a}. This could suggest that particle loss is
not an important concern. On the other hand, it has been suggested that the development 
of a power law requires a loss mechanism in addition to an energy drive \cite{drake13a}.
However, several recent simulations suggest that power-law spectra may still develop in 
the absence of a loss mechanism \cite{guo14a,sironi14a,li15a,werner15a}.
The set of conditions under which power-law spectra form in kinetic reconnection
simulations remains an open issue.

The simulations have a number of numerical limitations. These include comparatively
small spatial and temporal scales, and an artificial electron-to-proton mass ratio. 
However, as was discussed in Section \ref{sec:conditions},
the greater characteristic velocity of a realistically `light' electron facilitates
transport in the stochastic topology; the efficiency of Fermi acceleration does not 
directly dependent on the particle mass. The largest simulations show that three-dimensional
dynamics are increasingly important at larger scales; in contrast the diminishing frequency of island
mergers leads to suppression of further acceleration in two-dimensional systems.

The role of the guide field in magnetic reconnection has broad implications for 
reconnection-driven particle acceleration in astrophysical systems. Electron acceleration is most efficient
in the regime where both (a) the Fermi mechanism operates and (b) strong three-dimensional
transport exists. The former requires $b_g \lesssim 1$, the latter $b_g > 0$, suggesting that
reconnection with a magnetic field of the same order as the reconnecting component ($b_g \sim 1$) 
will yield the most efficient energetic electron production.
This result is especially relevant for solar flares, where the shear in the magnetic configuration
typically diminishes during the impulsive phase (see Fletcher et al. \cite{fletcher11a} 
and references therein). This corresponds to a transition from strong to weak guide field 
reconnection, and could explain why electron acceleration (as inferred from hard X-ray emissions)
is typically confined to the impulsive phase.

\begin{acknowledgments}
This work has been supported by NSF Grants
AGS1202330 and PHY1102479, and NASA grants
NNX11AQ93H, APL-975268, NNX08AV87G, NAS 5-
98033, and NNX08AO83G. J.T.D. acknowledges
support from the NASA LWS Jack Eddy Fellowship
administered by the University Corporation for Atmospheric
Research in Boulder, Colorado. Simulations were carried out at
the National Energy Research Scientific Computing Center.
\end{acknowledgments}

\clearpage
\clearpage

\clearpage

\clearpage

\clearpage

\clearpage
\clearpage

\clearpage
\clearpage

\clearpage
\clearpage

\clearpage
\clearpage

\clearpage

\clearpage

\bibliography{../paper}

\begin{thebibliography}{56}%
\makeatletter
\providecommand \@ifxundefined [1]{%
 \@ifx{#1\undefined}
}%
\providecommand \@ifnum [1]{%
 \ifnum #1\expandafter \@firstoftwo
 \else \expandafter \@secondoftwo
 \fi
}%
\providecommand \@ifx [1]{%
 \ifx #1\expandafter \@firstoftwo
 \else \expandafter \@secondoftwo
 \fi
}%
\providecommand \natexlab [1]{#1}%
\providecommand \enquote  [1]{``#1''}%
\providecommand \bibnamefont  [1]{#1}%
\providecommand \bibfnamefont [1]{#1}%
\providecommand \citenamefont [1]{#1}%
\providecommand \href@noop [0]{\@secondoftwo}%
\providecommand \href [0]{\begingroup \@sanitize@url \@href}%
\providecommand \@href[1]{\@@startlink{#1}\@@href}%
\providecommand \@@href[1]{\endgroup#1\@@endlink}%
\providecommand \@sanitize@url [0]{\catcode `\\12\catcode `\$12\catcode
  `\&12\catcode `\#12\catcode `\^12\catcode `\_12\catcode `\%12\relax}%
\providecommand \@@startlink[1]{}%
\providecommand \@@endlink[0]{}%
\providecommand \url  [0]{\begingroup\@sanitize@url \@url }%
\providecommand \@url [1]{\endgroup\@href {#1}{\urlprefix }}%
\providecommand \urlprefix  [0]{URL }%
\providecommand \Eprint [0]{\href }%
\providecommand \doibase [0]{http://dx.doi.org/}%
\providecommand \selectlanguage [0]{\@gobble}%
\providecommand \bibinfo  [0]{\@secondoftwo}%
\providecommand \bibfield  [0]{\@secondoftwo}%
\providecommand \translation [1]{[#1]}%
\providecommand \BibitemOpen [0]{}%
\providecommand \bibitemStop [0]{}%
\providecommand \bibitemNoStop [0]{.\EOS\space}%
\providecommand \EOS [0]{\spacefactor3000\relax}%
\providecommand \BibitemShut  [1]{\csname bibitem#1\endcsname}%
\let\auto@bib@innerbib\@empty
\bibitem [{\citenamefont {Lin}\ \emph {et~al.}(2003)\citenamefont {Lin},
  \citenamefont {Krucker}, \citenamefont {Hurford}, \citenamefont {Smith},
  \citenamefont {Hudson}, \citenamefont {Holman}, \citenamefont {Schwartz},
  \citenamefont {Dennis}, \citenamefont {Share}, \citenamefont {Murphy},
  \citenamefont {Emslie}, \citenamefont {Johns-Krull},\ and\ \citenamefont
  {Vilmer}}]{lin03a}%
  \BibitemOpen
  \bibfield  {author} {\bibinfo {author} {\bibfnamefont {R.~P.}\ \bibnamefont
  {Lin}}, \bibinfo {author} {\bibfnamefont {S.}~\bibnamefont {Krucker}},
  \bibinfo {author} {\bibfnamefont {G.~J.}\ \bibnamefont {Hurford}}, \bibinfo
  {author} {\bibfnamefont {D.~M.}\ \bibnamefont {Smith}}, \bibinfo {author}
  {\bibfnamefont {H.~S.}\ \bibnamefont {Hudson}}, \bibinfo {author}
  {\bibfnamefont {G.~D.}\ \bibnamefont {Holman}}, \bibinfo {author}
  {\bibfnamefont {R.~A.}\ \bibnamefont {Schwartz}}, \bibinfo {author}
  {\bibfnamefont {B.~R.}\ \bibnamefont {Dennis}}, \bibinfo {author}
  {\bibfnamefont {G.~H.}\ \bibnamefont {Share}}, \bibinfo {author}
  {\bibfnamefont {R.~J.}\ \bibnamefont {Murphy}}, \bibinfo {author}
  {\bibfnamefont {A.~G.}\ \bibnamefont {Emslie}}, \bibinfo {author}
  {\bibfnamefont {C.}~\bibnamefont {Johns-Krull}}, \ and\ \bibinfo {author}
  {\bibfnamefont {N.}~\bibnamefont {Vilmer}},\ }\href@noop {} {\bibfield
  {journal} {\bibinfo  {journal} {Ap. J.}\ }\textbf {\bibinfo {volume} {595}},\
  \bibinfo {pages} {L69} (\bibinfo {year} {2003})}\BibitemShut {NoStop}%
\bibitem [{\citenamefont {{\O}ieroset}\ \emph {et~al.}(2002)\citenamefont
  {{\O}ieroset}, \citenamefont {Lin}, \citenamefont {Phan}, \citenamefont
  {Larson},\ and\ \citenamefont {Bale}}]{oieroset02a}%
  \BibitemOpen
  \bibfield  {author} {\bibinfo {author} {\bibfnamefont {M.}~\bibnamefont
  {{\O}ieroset}}, \bibinfo {author} {\bibfnamefont {R.~P.}\ \bibnamefont
  {Lin}}, \bibinfo {author} {\bibfnamefont {T.~D.}\ \bibnamefont {Phan}},
  \bibinfo {author} {\bibfnamefont {D.~E.}\ \bibnamefont {Larson}}, \ and\
  \bibinfo {author} {\bibfnamefont {S.~D.}\ \bibnamefont {Bale}},\ }\href@noop
  {} {\bibfield  {journal} {\bibinfo  {journal} {Phys. Rev. Lett.}\ }\textbf
  {\bibinfo {volume} {89}},\ \bibinfo {pages} {195001} (\bibinfo {year}
  {2002})}\BibitemShut {NoStop}%
\bibitem [{\citenamefont {Krucker}\ \emph {et~al.}(2010)\citenamefont
  {Krucker}, \citenamefont {Hudson}, \citenamefont {Glesener}, \citenamefont
  {White}, \citenamefont {Masuda}, \citenamefont {Wuelser},\ and\ \citenamefont
  {Lin}}]{krucker10a}%
  \BibitemOpen
  \bibfield  {author} {\bibinfo {author} {\bibfnamefont {S.}~\bibnamefont
  {Krucker}}, \bibinfo {author} {\bibfnamefont {H.~S.}\ \bibnamefont {Hudson}},
  \bibinfo {author} {\bibfnamefont {L.}~\bibnamefont {Glesener}}, \bibinfo
  {author} {\bibfnamefont {S.~M.}\ \bibnamefont {White}}, \bibinfo {author}
  {\bibfnamefont {S.}~\bibnamefont {Masuda}}, \bibinfo {author} {\bibfnamefont
  {J.-P.}\ \bibnamefont {Wuelser}}, \ and\ \bibinfo {author} {\bibfnamefont
  {R.~P.}\ \bibnamefont {Lin}},\ }\href {\doibase 10.1088/0004-637X/714/2/1108}
  {\bibfield  {journal} {\bibinfo  {journal} {Ap. J.}\ }\textbf {\bibinfo
  {volume} {714}},\ \bibinfo {pages} {1108} (\bibinfo {year}
  {2010})}\BibitemShut {NoStop}%
\bibitem [{\citenamefont {Oka}\ \emph {et~al.}(2013)\citenamefont {Oka},
  \citenamefont {Ishikawa}, \citenamefont {Saint-Hilaire}, \citenamefont
  {Krucker},\ and\ \citenamefont {Lin}}]{oka13a}%
  \BibitemOpen
  \bibfield  {author} {\bibinfo {author} {\bibfnamefont {M.}~\bibnamefont
  {Oka}}, \bibinfo {author} {\bibfnamefont {S.}~\bibnamefont {Ishikawa}},
  \bibinfo {author} {\bibfnamefont {P.}~\bibnamefont {Saint-Hilaire}}, \bibinfo
  {author} {\bibfnamefont {S.}~\bibnamefont {Krucker}}, \ and\ \bibinfo
  {author} {\bibfnamefont {R.~P.}\ \bibnamefont {Lin}},\ }\href
  {http://stacks.iop.org/0004-637X/764/i=1/a=6} {\bibfield  {journal} {\bibinfo
   {journal} {The Astrophysical Journal}\ }\textbf {\bibinfo {volume} {764}},\
  \bibinfo {pages} {6} (\bibinfo {year} {2013})}\BibitemShut {NoStop}%
\bibitem [{\citenamefont {Drenkhahn}\ and\ \citenamefont
  {Spruit}(2002)}]{drenkhahn02a}%
  \BibitemOpen
  \bibfield  {author} {\bibinfo {author} {\bibfnamefont {G.}~\bibnamefont
  {Drenkhahn}}\ and\ \bibinfo {author} {\bibfnamefont {H.~C.}\ \bibnamefont
  {Spruit}},\ }\href {\doibase 10.1051/0004-6361:20020839} {\bibfield
  {journal} {\bibinfo  {journal} {Astronomy \& Astrophysics}\ }\textbf
  {\bibinfo {volume} {391}},\ \bibinfo {pages} {1141} (\bibinfo {year}
  {2002})}\BibitemShut {NoStop}%
\bibitem [{\citenamefont {Michel}(1994)}]{michel94a}%
  \BibitemOpen
  \bibfield  {author} {\bibinfo {author} {\bibfnamefont {F.~C.}\ \bibnamefont
  {Michel}},\ }\href@noop {} {\bibfield  {journal} {\bibinfo  {journal} {Ap.
  J.}\ }\textbf {\bibinfo {volume} {431}},\ \bibinfo {pages} {397} (\bibinfo
  {year} {1994})}\BibitemShut {NoStop}%
\bibitem [{\citenamefont {Tavani}\ \emph {et~al.}(2011)\citenamefont {Tavani},
  \citenamefont {Bulgarelli}, \citenamefont {Vittorini}, \citenamefont
  {Pellizzoni}, \citenamefont {Striani}, \citenamefont {Caraveo}, \citenamefont
  {Weisskopf}, \citenamefont {Tennant}, \citenamefont {Pucella}, \citenamefont
  {Trois}, \citenamefont {Costa}, \citenamefont {Evangelista}, \citenamefont
  {Pittori}, \citenamefont {Verrecchia}, \citenamefont {Del~Monte},
  \citenamefont {Campana}, \citenamefont {Pilia}, \citenamefont {De~Luca},
  \citenamefont {Donnarumma}, \citenamefont {Horns}, \citenamefont {Ferrigno},
  \citenamefont {Heinke}, \citenamefont {Trifoglio}, \citenamefont {Gianotti},
  \citenamefont {Vercellone}, \citenamefont {Argan}, \citenamefont
  {Barbiellini}, \citenamefont {Cattaneo}, \citenamefont {Chen}, \citenamefont
  {Contessi}, \citenamefont {D'Ammando}, \citenamefont {DeParis},
  \citenamefont {Di~Cocco}, \citenamefont {Di~Persio}, \citenamefont {Feroci},
  \citenamefont {Ferrari}, \citenamefont {Galli}, \citenamefont {Giuliani},
  \citenamefont {Giusti}, \citenamefont {Labanti}, \citenamefont {Lapshov},
  \citenamefont {Lazzarotto}, \citenamefont {Lipari}, \citenamefont {Longo},
  \citenamefont {Fuschino}, \citenamefont {Marisaldi}, \citenamefont
  {Mereghetti}, \citenamefont {Morelli}, \citenamefont {Moretti}, \citenamefont
  {Morselli}, \citenamefont {Pacciani}, \citenamefont {Perotti}, \citenamefont
  {Piano}, \citenamefont {Picozza}, \citenamefont {Prest}, \citenamefont
  {Rapisarda}, \citenamefont {Rappoldi}, \citenamefont {Rubini}, \citenamefont
  {Sabatini}, \citenamefont {Soffitta}, \citenamefont {Vallazza}, \citenamefont
  {Zambra}, \citenamefont {Zanello}, \citenamefont {Lucarelli}, \citenamefont
  {Santolamazza}, \citenamefont {Giommi}, \citenamefont {Salotti},\ and\
  \citenamefont {Bignami}}]{tavani11a}%
  \BibitemOpen
  \bibfield  {author} {\bibinfo {author} {\bibfnamefont {M.}~\bibnamefont
  {Tavani}}, \bibinfo {author} {\bibfnamefont {A.}~\bibnamefont {Bulgarelli}},
  \bibinfo {author} {\bibfnamefont {V.}~\bibnamefont {Vittorini}}, \bibinfo
  {author} {\bibfnamefont {A.}~\bibnamefont {Pellizzoni}}, \bibinfo {author}
  {\bibfnamefont {E.}~\bibnamefont {Striani}}, \bibinfo {author} {\bibfnamefont
  {P.}~\bibnamefont {Caraveo}}, \bibinfo {author} {\bibfnamefont {M.~C.}\
  \bibnamefont {Weisskopf}}, \bibinfo {author} {\bibfnamefont {A.}~\bibnamefont
  {Tennant}}, \bibinfo {author} {\bibfnamefont {G.}~\bibnamefont {Pucella}},
  \bibinfo {author} {\bibfnamefont {A.}~\bibnamefont {Trois}}, \bibinfo
  {author} {\bibfnamefont {E.}~\bibnamefont {Costa}}, \bibinfo {author}
  {\bibfnamefont {Y.}~\bibnamefont {Evangelista}}, \bibinfo {author}
  {\bibfnamefont {C.}~\bibnamefont {Pittori}}, \bibinfo {author} {\bibfnamefont
  {F.}~\bibnamefont {Verrecchia}}, \bibinfo {author} {\bibfnamefont
  {E.}~\bibnamefont {Del~Monte}}, \bibinfo {author} {\bibfnamefont
  {R.}~\bibnamefont {Campana}}, \bibinfo {author} {\bibfnamefont
  {M.}~\bibnamefont {Pilia}}, \bibinfo {author} {\bibfnamefont
  {A.}~\bibnamefont {De~Luca}}, \bibinfo {author} {\bibfnamefont
  {I.}~\bibnamefont {Donnarumma}}, \bibinfo {author} {\bibfnamefont
  {D.}~\bibnamefont {Horns}}, \bibinfo {author} {\bibfnamefont
  {C.}~\bibnamefont {Ferrigno}}, \bibinfo {author} {\bibfnamefont {C.~O.}\
  \bibnamefont {Heinke}}, \bibinfo {author} {\bibfnamefont {M.}~\bibnamefont
  {Trifoglio}}, \bibinfo {author} {\bibfnamefont {F.}~\bibnamefont {Gianotti}},
  \bibinfo {author} {\bibfnamefont {S.}~\bibnamefont {Vercellone}}, \bibinfo
  {author} {\bibfnamefont {A.}~\bibnamefont {Argan}}, \bibinfo {author}
  {\bibfnamefont {G.}~\bibnamefont {Barbiellini}}, \bibinfo {author}
  {\bibfnamefont {P.~W.}\ \bibnamefont {Cattaneo}}, \bibinfo {author}
  {\bibfnamefont {A.~W.}\ \bibnamefont {Chen}}, \bibinfo {author}
  {\bibfnamefont {T.}~\bibnamefont {Contessi}}, \bibinfo {author}
  {\bibfnamefont {F.}~\bibnamefont {D'Ammando}}, \bibinfo {author}
  {\bibfnamefont {G.}~\bibnamefont {DeParis}}, \bibinfo {author} {\bibfnamefont
  {G.}~\bibnamefont {Di~Cocco}}, \bibinfo {author} {\bibfnamefont
  {G.}~\bibnamefont {Di~Persio}}, \bibinfo {author} {\bibfnamefont
  {M.}~\bibnamefont {Feroci}}, \bibinfo {author} {\bibfnamefont
  {A.}~\bibnamefont {Ferrari}}, \bibinfo {author} {\bibfnamefont
  {M.}~\bibnamefont {Galli}}, \bibinfo {author} {\bibfnamefont
  {A.}~\bibnamefont {Giuliani}}, \bibinfo {author} {\bibfnamefont
  {M.}~\bibnamefont {Giusti}}, \bibinfo {author} {\bibfnamefont
  {C.}~\bibnamefont {Labanti}}, \bibinfo {author} {\bibfnamefont
  {I.}~\bibnamefont {Lapshov}}, \bibinfo {author} {\bibfnamefont
  {F.}~\bibnamefont {Lazzarotto}}, \bibinfo {author} {\bibfnamefont
  {P.}~\bibnamefont {Lipari}}, \bibinfo {author} {\bibfnamefont
  {F.}~\bibnamefont {Longo}}, \bibinfo {author} {\bibfnamefont
  {F.}~\bibnamefont {Fuschino}}, \bibinfo {author} {\bibfnamefont
  {M.}~\bibnamefont {Marisaldi}}, \bibinfo {author} {\bibfnamefont
  {S.}~\bibnamefont {Mereghetti}}, \bibinfo {author} {\bibfnamefont
  {E.}~\bibnamefont {Morelli}}, \bibinfo {author} {\bibfnamefont
  {E.}~\bibnamefont {Moretti}}, \bibinfo {author} {\bibfnamefont
  {A.}~\bibnamefont {Morselli}}, \bibinfo {author} {\bibfnamefont
  {L.}~\bibnamefont {Pacciani}}, \bibinfo {author} {\bibfnamefont
  {F.}~\bibnamefont {Perotti}}, \bibinfo {author} {\bibfnamefont
  {G.}~\bibnamefont {Piano}}, \bibinfo {author} {\bibfnamefont
  {P.}~\bibnamefont {Picozza}}, \bibinfo {author} {\bibfnamefont
  {M.}~\bibnamefont {Prest}}, \bibinfo {author} {\bibfnamefont
  {M.}~\bibnamefont {Rapisarda}}, \bibinfo {author} {\bibfnamefont
  {A.}~\bibnamefont {Rappoldi}}, \bibinfo {author} {\bibfnamefont
  {A.}~\bibnamefont {Rubini}}, \bibinfo {author} {\bibfnamefont
  {S.}~\bibnamefont {Sabatini}}, \bibinfo {author} {\bibfnamefont
  {P.}~\bibnamefont {Soffitta}}, \bibinfo {author} {\bibfnamefont
  {E.}~\bibnamefont {Vallazza}}, \bibinfo {author} {\bibfnamefont
  {A.}~\bibnamefont {Zambra}}, \bibinfo {author} {\bibfnamefont
  {D.}~\bibnamefont {Zanello}}, \bibinfo {author} {\bibfnamefont
  {F.}~\bibnamefont {Lucarelli}}, \bibinfo {author} {\bibfnamefont
  {P.}~\bibnamefont {Santolamazza}}, \bibinfo {author} {\bibfnamefont
  {P.}~\bibnamefont {Giommi}}, \bibinfo {author} {\bibfnamefont
  {L.}~\bibnamefont {Salotti}}, \ and\ \bibinfo {author} {\bibfnamefont
  {G.~F.}\ \bibnamefont {Bignami}},\ }\href {\doibase 10.1126/science.1200083}
  {\bibfield  {journal} {\bibinfo  {journal} {Science}\ }\textbf {\bibinfo
  {volume} {331}},\ \bibinfo {pages} {736} (\bibinfo {year}
  {2011})}\BibitemShut {NoStop}%
\bibitem [{\citenamefont {Abdo}\ \emph {et~al.}(2011)\citenamefont {Abdo},
  \citenamefont {Ackermann}, \citenamefont {Ajello}, \citenamefont {Allafort},
  \citenamefont {Baldini}, \citenamefont {Ballet}, \citenamefont {Barbiellini},
  \citenamefont {Bastieri}, \citenamefont {Bechtol}, \citenamefont
  {Bellazzini}, \citenamefont {Berenji}, \citenamefont {Blandford},
  \citenamefont {Bloom}, \citenamefont {Bonamente}, \citenamefont {Borgland},
  \citenamefont {Bouvier}, \citenamefont {Brandt}, \citenamefont {Bregeon},
  \citenamefont {Brez}, \citenamefont {Brigida}, \citenamefont {Bruel},
  \citenamefont {Buehler}, \citenamefont {Buson}, \citenamefont {Caliandro},
  \citenamefont {Cameron}, \citenamefont {Cannon}, \citenamefont {Caraveo},
  \citenamefont {Casandjian}, \citenamefont {Çelik}, \citenamefont {Charles},
  \citenamefont {Chekhtman}, \citenamefont {Cheung}, \citenamefont {Chiang},
  \citenamefont {Ciprini}, \citenamefont {Claus}, \citenamefont {Cohen-Tanugi},
  \citenamefont {Costamante}, \citenamefont {Cutini}, \citenamefont
  {D'Ammando}, \citenamefont {Dermer}, \citenamefont {de~Angelis},
  \citenamefont {de~Luca}, \citenamefont {de~Palma}, \citenamefont {Digel},
  \citenamefont {do~Couto~e Silva}, \citenamefont {Drell}, \citenamefont
  {Drlica-Wagner}, \citenamefont {Dubois}, \citenamefont {Dumora},
  \citenamefont {Favuzzi}, \citenamefont {Fegan}, \citenamefont {Ferrara},
  \citenamefont {Focke}, \citenamefont {Fortin}, \citenamefont {Frailis},
  \citenamefont {Fukazawa}, \citenamefont {Funk}, \citenamefont {Fusco},
  \citenamefont {Gargano}, \citenamefont {Gasparrini}, \citenamefont {Gehrels},
  \citenamefont {Germani}, \citenamefont {Giglietto}, \citenamefont {Giordano},
  \citenamefont {Giroletti}, \citenamefont {Glanzman}, \citenamefont {Godfrey},
  \citenamefont {Grenier}, \citenamefont {Grondin}, \citenamefont {Grove},
  \citenamefont {Guiriec}, \citenamefont {Hadasch}, \citenamefont {Hanabata},
  \citenamefont {Harding}, \citenamefont {Hayashi}, \citenamefont {Hayashida},
  \citenamefont {Hays}, \citenamefont {Horan}, \citenamefont {Itoh},
  \citenamefont {Jóhannesson}, \citenamefont {Johnson}, \citenamefont
  {Johnson}, \citenamefont {Khangulyan}, \citenamefont {Kamae}, \citenamefont
  {Katagiri}, \citenamefont {Kataoka}, \citenamefont {Kerr}, \citenamefont
  {Knödlseder}, \citenamefont {Kuss}, \citenamefont {Lande}, \citenamefont
  {Latronico}, \citenamefont {Lee}, \citenamefont {Lemoine-Goumard},
  \citenamefont {Longo}, \citenamefont {Loparco}, \citenamefont {Lubrano},
  \citenamefont {Madejski}, \citenamefont {Makeev}, \citenamefont {Marelli},
  \citenamefont {Mazziotta}, \citenamefont {McEnery}, \citenamefont
  {Michelson}, \citenamefont {Mitthumsiri}, \citenamefont {Mizuno},
  \citenamefont {Moiseev}, \citenamefont {Monte}, \citenamefont {Monzani},
  \citenamefont {Morselli}, \citenamefont {Moskalenko}, \citenamefont {Murgia},
  \citenamefont {Nakamori}, \citenamefont {Naumann-Godo}, \citenamefont
  {Nolan}, \citenamefont {Norris}, \citenamefont {Nuss}, \citenamefont
  {Ohsugi}, \citenamefont {Okumura}, \citenamefont {Omodei}, \citenamefont
  {Ormes}, \citenamefont {Ozaki}, \citenamefont {Paneque}, \citenamefont
  {Parent}, \citenamefont {Pelassa}, \citenamefont {Pepe}, \citenamefont
  {Pesce-Rollins}, \citenamefont {Pierbattista}, \citenamefont {Piron},
  \citenamefont {Porter}, \citenamefont {Rainò}, \citenamefont {Rando},
  \citenamefont {Ray}, \citenamefont {Razzano}, \citenamefont {Reimer},
  \citenamefont {Reimer}, \citenamefont {Reposeur}, \citenamefont {Ritz},
  \citenamefont {Romani}, \citenamefont {Sadrozinski}, \citenamefont {Sanchez},
  \citenamefont {Parkinson}, \citenamefont {Scargle}, \citenamefont {Schalk},
  \citenamefont {Sgrò}, \citenamefont {Siskind}, \citenamefont {Smith},
  \citenamefont {Spandre}, \citenamefont {Spinelli}, \citenamefont {Strickman},
  \citenamefont {Suson}, \citenamefont {Takahashi}, \citenamefont {Takahashi},
  \citenamefont {Tanaka}, \citenamefont {Thayer}, \citenamefont {Thompson},
  \citenamefont {Tibaldo}, \citenamefont {Torres}, \citenamefont {Tosti},
  \citenamefont {Tramacere}, \citenamefont {Troja}, \citenamefont {Uchiyama},
  \citenamefont {Vandenbroucke}, \citenamefont {Vasileiou}, \citenamefont
  {Vianello}, \citenamefont {Vitale}, \citenamefont {Wang}, \citenamefont
  {Wood}, \citenamefont {Yang},\ and\ \citenamefont {Ziegler}}]{abdo11a}%
  \BibitemOpen
  \bibfield  {author} {\bibinfo {author} {\bibfnamefont {A.~A.}\ \bibnamefont
  {Abdo}}, \bibinfo {author} {\bibfnamefont {M.}~\bibnamefont {Ackermann}},
  \bibinfo {author} {\bibfnamefont {M.}~\bibnamefont {Ajello}}, \bibinfo
  {author} {\bibfnamefont {A.}~\bibnamefont {Allafort}}, \bibinfo {author}
  {\bibfnamefont {L.}~\bibnamefont {Baldini}}, \bibinfo {author} {\bibfnamefont
  {J.}~\bibnamefont {Ballet}}, \bibinfo {author} {\bibfnamefont
  {G.}~\bibnamefont {Barbiellini}}, \bibinfo {author} {\bibfnamefont
  {D.}~\bibnamefont {Bastieri}}, \bibinfo {author} {\bibfnamefont
  {K.}~\bibnamefont {Bechtol}}, \bibinfo {author} {\bibfnamefont
  {R.}~\bibnamefont {Bellazzini}}, \bibinfo {author} {\bibfnamefont
  {B.}~\bibnamefont {Berenji}}, \bibinfo {author} {\bibfnamefont {R.~D.}\
  \bibnamefont {Blandford}}, \bibinfo {author} {\bibfnamefont {E.~D.}\
  \bibnamefont {Bloom}}, \bibinfo {author} {\bibfnamefont {E.}~\bibnamefont
  {Bonamente}}, \bibinfo {author} {\bibfnamefont {A.~W.}\ \bibnamefont
  {Borgland}}, \bibinfo {author} {\bibfnamefont {A.}~\bibnamefont {Bouvier}},
  \bibinfo {author} {\bibfnamefont {T.~J.}\ \bibnamefont {Brandt}}, \bibinfo
  {author} {\bibfnamefont {J.}~\bibnamefont {Bregeon}}, \bibinfo {author}
  {\bibfnamefont {A.}~\bibnamefont {Brez}}, \bibinfo {author} {\bibfnamefont
  {M.}~\bibnamefont {Brigida}}, \bibinfo {author} {\bibfnamefont
  {P.}~\bibnamefont {Bruel}}, \bibinfo {author} {\bibfnamefont
  {R.}~\bibnamefont {Buehler}}, \bibinfo {author} {\bibfnamefont
  {S.}~\bibnamefont {Buson}}, \bibinfo {author} {\bibfnamefont {G.~A.}\
  \bibnamefont {Caliandro}}, \bibinfo {author} {\bibfnamefont {R.~A.}\
  \bibnamefont {Cameron}}, \bibinfo {author} {\bibfnamefont {A.}~\bibnamefont
  {Cannon}}, \bibinfo {author} {\bibfnamefont {P.~A.}\ \bibnamefont {Caraveo}},
  \bibinfo {author} {\bibfnamefont {J.~M.}\ \bibnamefont {Casandjian}},
  \bibinfo {author} {\bibfnamefont {Ã.}~\bibnamefont {Çelik}}, \bibinfo
  {author} {\bibfnamefont {E.}~\bibnamefont {Charles}}, \bibinfo {author}
  {\bibfnamefont {A.}~\bibnamefont {Chekhtman}}, \bibinfo {author}
  {\bibfnamefont {C.~C.}\ \bibnamefont {Cheung}}, \bibinfo {author}
  {\bibfnamefont {J.}~\bibnamefont {Chiang}}, \bibinfo {author} {\bibfnamefont
  {S.}~\bibnamefont {Ciprini}}, \bibinfo {author} {\bibfnamefont
  {R.}~\bibnamefont {Claus}}, \bibinfo {author} {\bibfnamefont
  {J.}~\bibnamefont {Cohen-Tanugi}}, \bibinfo {author} {\bibfnamefont
  {L.}~\bibnamefont {Costamante}}, \bibinfo {author} {\bibfnamefont
  {S.}~\bibnamefont {Cutini}}, \bibinfo {author} {\bibfnamefont
  {F.}~\bibnamefont {D'Ammando}}, \bibinfo {author} {\bibfnamefont {C.~D.}\
  \bibnamefont {Dermer}}, \bibinfo {author} {\bibfnamefont {A.}~\bibnamefont
  {de~Angelis}}, \bibinfo {author} {\bibfnamefont {A.}~\bibnamefont {de~Luca}},
  \bibinfo {author} {\bibfnamefont {F.}~\bibnamefont {de~Palma}}, \bibinfo
  {author} {\bibfnamefont {S.~W.}\ \bibnamefont {Digel}}, \bibinfo {author}
  {\bibfnamefont {E.}~\bibnamefont {do~Couto~e Silva}}, \bibinfo {author}
  {\bibfnamefont {P.~S.}\ \bibnamefont {Drell}}, \bibinfo {author}
  {\bibfnamefont {A.}~\bibnamefont {Drlica-Wagner}}, \bibinfo {author}
  {\bibfnamefont {R.}~\bibnamefont {Dubois}}, \bibinfo {author} {\bibfnamefont
  {D.}~\bibnamefont {Dumora}}, \bibinfo {author} {\bibfnamefont
  {C.}~\bibnamefont {Favuzzi}}, \bibinfo {author} {\bibfnamefont {S.~J.}\
  \bibnamefont {Fegan}}, \bibinfo {author} {\bibfnamefont {E.~C.}\ \bibnamefont
  {Ferrara}}, \bibinfo {author} {\bibfnamefont {W.~B.}\ \bibnamefont {Focke}},
  \bibinfo {author} {\bibfnamefont {P.}~\bibnamefont {Fortin}}, \bibinfo
  {author} {\bibfnamefont {M.}~\bibnamefont {Frailis}}, \bibinfo {author}
  {\bibfnamefont {Y.}~\bibnamefont {Fukazawa}}, \bibinfo {author}
  {\bibfnamefont {S.}~\bibnamefont {Funk}}, \bibinfo {author} {\bibfnamefont
  {P.}~\bibnamefont {Fusco}}, \bibinfo {author} {\bibfnamefont
  {F.}~\bibnamefont {Gargano}}, \bibinfo {author} {\bibfnamefont
  {D.}~\bibnamefont {Gasparrini}}, \bibinfo {author} {\bibfnamefont
  {N.}~\bibnamefont {Gehrels}}, \bibinfo {author} {\bibfnamefont
  {S.}~\bibnamefont {Germani}}, \bibinfo {author} {\bibfnamefont
  {N.}~\bibnamefont {Giglietto}}, \bibinfo {author} {\bibfnamefont
  {F.}~\bibnamefont {Giordano}}, \bibinfo {author} {\bibfnamefont
  {M.}~\bibnamefont {Giroletti}}, \bibinfo {author} {\bibfnamefont
  {T.}~\bibnamefont {Glanzman}}, \bibinfo {author} {\bibfnamefont
  {G.}~\bibnamefont {Godfrey}}, \bibinfo {author} {\bibfnamefont {I.~A.}\
  \bibnamefont {Grenier}}, \bibinfo {author} {\bibfnamefont {M.-H.}\
  \bibnamefont {Grondin}}, \bibinfo {author} {\bibfnamefont {J.~E.}\
  \bibnamefont {Grove}}, \bibinfo {author} {\bibfnamefont {S.}~\bibnamefont
  {Guiriec}}, \bibinfo {author} {\bibfnamefont {D.}~\bibnamefont {Hadasch}},
  \bibinfo {author} {\bibfnamefont {Y.}~\bibnamefont {Hanabata}}, \bibinfo
  {author} {\bibfnamefont {A.~K.}\ \bibnamefont {Harding}}, \bibinfo {author}
  {\bibfnamefont {K.}~\bibnamefont {Hayashi}}, \bibinfo {author} {\bibfnamefont
  {M.}~\bibnamefont {Hayashida}}, \bibinfo {author} {\bibfnamefont
  {E.}~\bibnamefont {Hays}}, \bibinfo {author} {\bibfnamefont {D.}~\bibnamefont
  {Horan}}, \bibinfo {author} {\bibfnamefont {R.}~\bibnamefont {Itoh}},
  \bibinfo {author} {\bibfnamefont {G.}~\bibnamefont {Jóhannesson}}, \bibinfo
  {author} {\bibfnamefont {A.~S.}\ \bibnamefont {Johnson}}, \bibinfo {author}
  {\bibfnamefont {T.~J.}\ \bibnamefont {Johnson}}, \bibinfo {author}
  {\bibfnamefont {D.}~\bibnamefont {Khangulyan}}, \bibinfo {author}
  {\bibfnamefont {T.}~\bibnamefont {Kamae}}, \bibinfo {author} {\bibfnamefont
  {H.}~\bibnamefont {Katagiri}}, \bibinfo {author} {\bibfnamefont
  {J.}~\bibnamefont {Kataoka}}, \bibinfo {author} {\bibfnamefont
  {M.}~\bibnamefont {Kerr}}, \bibinfo {author} {\bibfnamefont {J.}~\bibnamefont
  {Knödlseder}}, \bibinfo {author} {\bibfnamefont {M.}~\bibnamefont {Kuss}},
  \bibinfo {author} {\bibfnamefont {J.}~\bibnamefont {Lande}}, \bibinfo
  {author} {\bibfnamefont {L.}~\bibnamefont {Latronico}}, \bibinfo {author}
  {\bibfnamefont {S.-H.}\ \bibnamefont {Lee}}, \bibinfo {author} {\bibfnamefont
  {M.}~\bibnamefont {Lemoine-Goumard}}, \bibinfo {author} {\bibfnamefont
  {F.}~\bibnamefont {Longo}}, \bibinfo {author} {\bibfnamefont
  {F.}~\bibnamefont {Loparco}}, \bibinfo {author} {\bibfnamefont
  {P.}~\bibnamefont {Lubrano}}, \bibinfo {author} {\bibfnamefont {G.~M.}\
  \bibnamefont {Madejski}}, \bibinfo {author} {\bibfnamefont {A.}~\bibnamefont
  {Makeev}}, \bibinfo {author} {\bibfnamefont {M.}~\bibnamefont {Marelli}},
  \bibinfo {author} {\bibfnamefont {M.~N.}\ \bibnamefont {Mazziotta}}, \bibinfo
  {author} {\bibfnamefont {J.~E.}\ \bibnamefont {McEnery}}, \bibinfo {author}
  {\bibfnamefont {P.~F.}\ \bibnamefont {Michelson}}, \bibinfo {author}
  {\bibfnamefont {W.}~\bibnamefont {Mitthumsiri}}, \bibinfo {author}
  {\bibfnamefont {T.}~\bibnamefont {Mizuno}}, \bibinfo {author} {\bibfnamefont
  {A.~A.}\ \bibnamefont {Moiseev}}, \bibinfo {author} {\bibfnamefont
  {C.}~\bibnamefont {Monte}}, \bibinfo {author} {\bibfnamefont {M.~E.}\
  \bibnamefont {Monzani}}, \bibinfo {author} {\bibfnamefont {A.}~\bibnamefont
  {Morselli}}, \bibinfo {author} {\bibfnamefont {I.~V.}\ \bibnamefont
  {Moskalenko}}, \bibinfo {author} {\bibfnamefont {S.}~\bibnamefont {Murgia}},
  \bibinfo {author} {\bibfnamefont {T.}~\bibnamefont {Nakamori}}, \bibinfo
  {author} {\bibfnamefont {M.}~\bibnamefont {Naumann-Godo}}, \bibinfo {author}
  {\bibfnamefont {P.~L.}\ \bibnamefont {Nolan}}, \bibinfo {author}
  {\bibfnamefont {J.~P.}\ \bibnamefont {Norris}}, \bibinfo {author}
  {\bibfnamefont {E.}~\bibnamefont {Nuss}}, \bibinfo {author} {\bibfnamefont
  {T.}~\bibnamefont {Ohsugi}}, \bibinfo {author} {\bibfnamefont
  {A.}~\bibnamefont {Okumura}}, \bibinfo {author} {\bibfnamefont
  {N.}~\bibnamefont {Omodei}}, \bibinfo {author} {\bibfnamefont {J.~F.}\
  \bibnamefont {Ormes}}, \bibinfo {author} {\bibfnamefont {M.}~\bibnamefont
  {Ozaki}}, \bibinfo {author} {\bibfnamefont {D.}~\bibnamefont {Paneque}},
  \bibinfo {author} {\bibfnamefont {D.}~\bibnamefont {Parent}}, \bibinfo
  {author} {\bibfnamefont {V.}~\bibnamefont {Pelassa}}, \bibinfo {author}
  {\bibfnamefont {M.}~\bibnamefont {Pepe}}, \bibinfo {author} {\bibfnamefont
  {M.}~\bibnamefont {Pesce-Rollins}}, \bibinfo {author} {\bibfnamefont
  {M.}~\bibnamefont {Pierbattista}}, \bibinfo {author} {\bibfnamefont
  {F.}~\bibnamefont {Piron}}, \bibinfo {author} {\bibfnamefont {T.~A.}\
  \bibnamefont {Porter}}, \bibinfo {author} {\bibfnamefont {S.}~\bibnamefont
  {Rainò}}, \bibinfo {author} {\bibfnamefont {R.}~\bibnamefont {Rando}},
  \bibinfo {author} {\bibfnamefont {P.~S.}\ \bibnamefont {Ray}}, \bibinfo
  {author} {\bibfnamefont {M.}~\bibnamefont {Razzano}}, \bibinfo {author}
  {\bibfnamefont {A.}~\bibnamefont {Reimer}}, \bibinfo {author} {\bibfnamefont
  {O.}~\bibnamefont {Reimer}}, \bibinfo {author} {\bibfnamefont
  {T.}~\bibnamefont {Reposeur}}, \bibinfo {author} {\bibfnamefont
  {S.}~\bibnamefont {Ritz}}, \bibinfo {author} {\bibfnamefont {R.~W.}\
  \bibnamefont {Romani}}, \bibinfo {author} {\bibfnamefont {H.~F.-W.}\
  \bibnamefont {Sadrozinski}}, \bibinfo {author} {\bibfnamefont
  {D.}~\bibnamefont {Sanchez}}, \bibinfo {author} {\bibfnamefont {P.~M.~S.}\
  \bibnamefont {Parkinson}}, \bibinfo {author} {\bibfnamefont {J.~D.}\
  \bibnamefont {Scargle}}, \bibinfo {author} {\bibfnamefont {T.~L.}\
  \bibnamefont {Schalk}}, \bibinfo {author} {\bibfnamefont {C.}~\bibnamefont
  {Sgrò}}, \bibinfo {author} {\bibfnamefont {E.~J.}\ \bibnamefont {Siskind}},
  \bibinfo {author} {\bibfnamefont {P.~D.}\ \bibnamefont {Smith}}, \bibinfo
  {author} {\bibfnamefont {G.}~\bibnamefont {Spandre}}, \bibinfo {author}
  {\bibfnamefont {P.}~\bibnamefont {Spinelli}}, \bibinfo {author}
  {\bibfnamefont {M.~S.}\ \bibnamefont {Strickman}}, \bibinfo {author}
  {\bibfnamefont {D.~J.}\ \bibnamefont {Suson}}, \bibinfo {author}
  {\bibfnamefont {H.}~\bibnamefont {Takahashi}}, \bibinfo {author}
  {\bibfnamefont {T.}~\bibnamefont {Takahashi}}, \bibinfo {author}
  {\bibfnamefont {T.}~\bibnamefont {Tanaka}}, \bibinfo {author} {\bibfnamefont
  {J.~B.}\ \bibnamefont {Thayer}}, \bibinfo {author} {\bibfnamefont {D.~J.}\
  \bibnamefont {Thompson}}, \bibinfo {author} {\bibfnamefont {L.}~\bibnamefont
  {Tibaldo}}, \bibinfo {author} {\bibfnamefont {D.~F.}\ \bibnamefont {Torres}},
  \bibinfo {author} {\bibfnamefont {G.}~\bibnamefont {Tosti}}, \bibinfo
  {author} {\bibfnamefont {A.}~\bibnamefont {Tramacere}}, \bibinfo {author}
  {\bibfnamefont {E.}~\bibnamefont {Troja}}, \bibinfo {author} {\bibfnamefont
  {Y.}~\bibnamefont {Uchiyama}}, \bibinfo {author} {\bibfnamefont
  {J.}~\bibnamefont {Vandenbroucke}}, \bibinfo {author} {\bibfnamefont
  {V.}~\bibnamefont {Vasileiou}}, \bibinfo {author} {\bibfnamefont
  {G.}~\bibnamefont {Vianello}}, \bibinfo {author} {\bibfnamefont
  {V.}~\bibnamefont {Vitale}}, \bibinfo {author} {\bibfnamefont
  {P.}~\bibnamefont {Wang}}, \bibinfo {author} {\bibfnamefont {K.~S.}\
  \bibnamefont {Wood}}, \bibinfo {author} {\bibfnamefont {Z.}~\bibnamefont
  {Yang}}, \ and\ \bibinfo {author} {\bibfnamefont {M.}~\bibnamefont
  {Ziegler}},\ }\href {\doibase 10.1126/science.1199705} {\bibfield  {journal}
  {\bibinfo  {journal} {Science}\ }\textbf {\bibinfo {volume} {331}},\ \bibinfo
  {pages} {739} (\bibinfo {year} {2011})}\BibitemShut {NoStop}%
\bibitem [{\citenamefont {Hoshino}\ \emph {et~al.}(2001)\citenamefont
  {Hoshino}, \citenamefont {Mukai}, \citenamefont {Terasawa},\ and\
  \citenamefont {Shinohara}}]{hoshino01a}%
  \BibitemOpen
  \bibfield  {author} {\bibinfo {author} {\bibfnamefont {M.}~\bibnamefont
  {Hoshino}}, \bibinfo {author} {\bibfnamefont {T.}~\bibnamefont {Mukai}},
  \bibinfo {author} {\bibfnamefont {T.}~\bibnamefont {Terasawa}}, \ and\
  \bibinfo {author} {\bibfnamefont {I.}~\bibnamefont {Shinohara}},\ }\href@noop
  {} {\bibfield  {journal} {\bibinfo  {journal} {J. Geophys. Res.}\ }\textbf
  {\bibinfo {volume} {106}},\ \bibinfo {pages} {25,979} (\bibinfo {year}
  {2001})}\BibitemShut {NoStop}%
\bibitem [{\citenamefont {Zenitani}\ and\ \citenamefont
  {Hoshino}(2001)}]{zenitani01a}%
  \BibitemOpen
  \bibfield  {author} {\bibinfo {author} {\bibfnamefont {S.}~\bibnamefont
  {Zenitani}}\ and\ \bibinfo {author} {\bibfnamefont {M.}~\bibnamefont
  {Hoshino}},\ }\href@noop {} {\bibfield  {journal} {\bibinfo  {journal} {Ap.
  J. Lett.}\ }\textbf {\bibinfo {volume} {562}},\ \bibinfo {pages} {L63}
  (\bibinfo {year} {2001})}\BibitemShut {NoStop}%
\bibitem [{\citenamefont {Drake}\ \emph {et~al.}(2005)\citenamefont {Drake},
  \citenamefont {Shay}, \citenamefont {Thongthai},\ and\ \citenamefont
  {Swisdak}}]{drake05a}%
  \BibitemOpen
  \bibfield  {author} {\bibinfo {author} {\bibfnamefont {J.~F.}\ \bibnamefont
  {Drake}}, \bibinfo {author} {\bibfnamefont {M.~A.}\ \bibnamefont {Shay}},
  \bibinfo {author} {\bibfnamefont {W.}~\bibnamefont {Thongthai}}, \ and\
  \bibinfo {author} {\bibfnamefont {M.}~\bibnamefont {Swisdak}},\ }\href@noop
  {} {\bibfield  {journal} {\bibinfo  {journal} {Phys. Rev. Lett.}\ }\textbf
  {\bibinfo {volume} {94}},\ \bibinfo {pages} {095001} (\bibinfo {year}
  {2005})}\BibitemShut {NoStop}%
\bibitem [{\citenamefont {Pritchett}(2006)}]{pritchett06a}%
  \BibitemOpen
  \bibfield  {author} {\bibinfo {author} {\bibfnamefont {P.~L.}\ \bibnamefont
  {Pritchett}},\ }\href {\doibase 10.1029/2006JA011793} {\bibfield  {journal}
  {\bibinfo  {journal} {J. Geophys. Res.}\ }\textbf {\bibinfo {volume} {111}},\
  \bibinfo {eid} {A10212} (\bibinfo {year} {2006}),\
  10.1029/2006JA011793}\BibitemShut {NoStop}%
\bibitem [{\citenamefont {Egedal}\ \emph {et~al.}(2009)\citenamefont {Egedal},
  \citenamefont {Daughton}, \citenamefont {Drake}, \citenamefont {Katz},\ and\
  \citenamefont {L{\^e}}}]{egedal09a}%
  \BibitemOpen
  \bibfield  {author} {\bibinfo {author} {\bibfnamefont {J.}~\bibnamefont
  {Egedal}}, \bibinfo {author} {\bibfnamefont {W.}~\bibnamefont {Daughton}},
  \bibinfo {author} {\bibfnamefont {J.~F.}\ \bibnamefont {Drake}}, \bibinfo
  {author} {\bibfnamefont {N.}~\bibnamefont {Katz}}, \ and\ \bibinfo {author}
  {\bibfnamefont {A.}~\bibnamefont {L{\^e}}},\ }\href {\doibase
  10.1063/1.3130732} {\bibfield  {journal} {\bibinfo  {journal} {Phys.
  Plasmas}\ }\textbf {\bibinfo {volume} {16}},\ \bibinfo {eid} {050701}
  (\bibinfo {year} {2009}),\ 10.1063/1.3130732}\BibitemShut {NoStop}%
\bibitem [{\citenamefont {Oka}\ \emph {et~al.}(2010)\citenamefont {Oka},
  \citenamefont {Phan}, \citenamefont {Krucker}, \citenamefont {Fujimoto},\
  and\ \citenamefont {Shinohara}}]{oka10a}%
  \BibitemOpen
  \bibfield  {author} {\bibinfo {author} {\bibfnamefont {M.}~\bibnamefont
  {Oka}}, \bibinfo {author} {\bibfnamefont {T.-D.}\ \bibnamefont {Phan}},
  \bibinfo {author} {\bibfnamefont {S.}~\bibnamefont {Krucker}}, \bibinfo
  {author} {\bibfnamefont {M.}~\bibnamefont {Fujimoto}}, \ and\ \bibinfo
  {author} {\bibfnamefont {I.}~\bibnamefont {Shinohara}},\ }\href {\doibase
  10.1088/0004-637X/714/1/915} {\bibfield  {journal} {\bibinfo  {journal} {Ap.
  J.}\ }\textbf {\bibinfo {volume} {714}},\ \bibinfo {pages} {915} (\bibinfo
  {year} {2010})}\BibitemShut {NoStop}%
\bibitem [{\citenamefont {Hoshino}(2012)}]{hoshino12a}%
  \BibitemOpen
  \bibfield  {author} {\bibinfo {author} {\bibfnamefont {M.}~\bibnamefont
  {Hoshino}},\ }\href {\doibase 10.1103/PhysRevLett.108.135003} {\bibfield
  {journal} {\bibinfo  {journal} {Phys. Rev. Lett.}\ }\textbf {\bibinfo
  {volume} {108}},\ \bibinfo {pages} {135003} (\bibinfo {year}
  {2012})}\BibitemShut {NoStop}%
\bibitem [{\citenamefont {Litvinenko}(1996)}]{litvinenko96a}%
  \BibitemOpen
  \bibfield  {author} {\bibinfo {author} {\bibfnamefont {Y.~E.}\ \bibnamefont
  {Litvinenko}},\ }\href@noop {} {\bibfield  {journal} {\bibinfo  {journal}
  {Ap. J.}\ }\textbf {\bibinfo {volume} {462}},\ \bibinfo {pages} {997}
  (\bibinfo {year} {1996})}\BibitemShut {NoStop}%
\bibitem [{\citenamefont {Uzdensky}, \citenamefont {Cerutti},\ and\
  \citenamefont {Begelman}(2011)}]{uzdensky11a}%
  \BibitemOpen
  \bibfield  {author} {\bibinfo {author} {\bibfnamefont {D.~A.}\ \bibnamefont
  {Uzdensky}}, \bibinfo {author} {\bibfnamefont {B.}~\bibnamefont {Cerutti}}, \
  and\ \bibinfo {author} {\bibfnamefont {M.~C.}\ \bibnamefont {Begelman}},\
  }\href@noop {} {\bibfield  {journal} {\bibinfo  {journal} {The Astrophysical
  Journal Letters}\ }\textbf {\bibinfo {volume} {737}},\ \bibinfo {pages} {L40}
  (\bibinfo {year} {2011})}\BibitemShut {NoStop}%
\bibitem [{\citenamefont {Egedal}, \citenamefont {Daughton},\ and\
  \citenamefont {L{\^e}}(2012)}]{egedal12a}%
  \BibitemOpen
  \bibfield  {author} {\bibinfo {author} {\bibfnamefont {J.}~\bibnamefont
  {Egedal}}, \bibinfo {author} {\bibfnamefont {W.}~\bibnamefont {Daughton}}, \
  and\ \bibinfo {author} {\bibfnamefont {A.}~\bibnamefont {L{\^e}}},\ }\href
  {\doibase 10.1038/nphys2249} {\bibfield  {journal} {\bibinfo  {journal}
  {Nature Phys.}\ }\textbf {\bibinfo {volume} {8}},\ \bibinfo {pages} {321}
  (\bibinfo {year} {2012})}\BibitemShut {NoStop}%
\bibitem [{\citenamefont {Dahlin}, \citenamefont {Drake},\ and\ \citenamefont
  {Swisdak}(2015)}]{dahlin15a}%
  \BibitemOpen
  \bibfield  {author} {\bibinfo {author} {\bibfnamefont {J.~T.}\ \bibnamefont
  {Dahlin}}, \bibinfo {author} {\bibfnamefont {J.~F.}\ \bibnamefont {Drake}}, \
  and\ \bibinfo {author} {\bibfnamefont {M.}~\bibnamefont {Swisdak}},\ }\href
  {\doibase http://dx.doi.org/10.1063/1.4933212} {\bibfield  {journal}
  {\bibinfo  {journal} {Physics of Plasmas}\ }\textbf {\bibinfo {volume}
  {22}},\ \bibinfo {eid} {100704} (\bibinfo {year} {2015}),\
  http://dx.doi.org/10.1063/1.4933212}\BibitemShut {NoStop}%
\bibitem [{\citenamefont {Drake}\ \emph {et~al.}(2006)\citenamefont {Drake},
  \citenamefont {Swisdak}, \citenamefont {Che},\ and\ \citenamefont
  {Shay}}]{drake06a}%
  \BibitemOpen
  \bibfield  {author} {\bibinfo {author} {\bibfnamefont {J.~F.}\ \bibnamefont
  {Drake}}, \bibinfo {author} {\bibfnamefont {M.}~\bibnamefont {Swisdak}},
  \bibinfo {author} {\bibfnamefont {H.}~\bibnamefont {Che}}, \ and\ \bibinfo
  {author} {\bibfnamefont {M.~A.}\ \bibnamefont {Shay}},\ }\href {\doibase
  10.1038/nature05116} {\bibfield  {journal} {\bibinfo  {journal} {Nature}\
  }\textbf {\bibinfo {volume} {443}},\ \bibinfo {pages} {553} (\bibinfo {year}
  {2006})}\BibitemShut {NoStop}%
\bibitem [{\citenamefont {Drake}\ \emph {et~al.}(2010)\citenamefont {Drake},
  \citenamefont {Opher}, \citenamefont {Swisdak},\ and\ \citenamefont
  {Chamoun}}]{drake10a}%
  \BibitemOpen
  \bibfield  {author} {\bibinfo {author} {\bibfnamefont {J.~F.}\ \bibnamefont
  {Drake}}, \bibinfo {author} {\bibfnamefont {M.}~\bibnamefont {Opher}},
  \bibinfo {author} {\bibfnamefont {M.}~\bibnamefont {Swisdak}}, \ and\
  \bibinfo {author} {\bibfnamefont {J.~N.}\ \bibnamefont {Chamoun}},\ }\href
  {\doibase 10.1088/0004-637X/709/2/963} {\bibfield  {journal} {\bibinfo
  {journal} {Ap. J.}\ }\textbf {\bibinfo {volume} {709}},\ \bibinfo {pages}
  {963} (\bibinfo {year} {2010})}\BibitemShut {NoStop}%
\bibitem [{\citenamefont {Drake}, \citenamefont {Swisdak},\ and\ \citenamefont
  {Fermo}(2013)}]{drake13a}%
  \BibitemOpen
  \bibfield  {author} {\bibinfo {author} {\bibfnamefont {J.~F.}\ \bibnamefont
  {Drake}}, \bibinfo {author} {\bibfnamefont {M.}~\bibnamefont {Swisdak}}, \
  and\ \bibinfo {author} {\bibfnamefont {R.}~\bibnamefont {Fermo}},\ }\href
  {http://stacks.iop.org/2041-8205/763/i=1/a=L5} {\bibfield  {journal}
  {\bibinfo  {journal} {The Astrophysical Journal Letters}\ }\textbf {\bibinfo
  {volume} {763}},\ \bibinfo {pages} {L5} (\bibinfo {year} {2013})}\BibitemShut
  {NoStop}%
\bibitem [{\citenamefont {Dahlin}, \citenamefont {Drake},\ and\ \citenamefont
  {Swisdak}(2016)}]{dahlin16a}%
  \BibitemOpen
  \bibfield  {author} {\bibinfo {author} {\bibfnamefont {J.~T.}\ \bibnamefont
  {Dahlin}}, \bibinfo {author} {\bibfnamefont {J.~F.}\ \bibnamefont {Drake}}, \
  and\ \bibinfo {author} {\bibfnamefont {M.}~\bibnamefont {Swisdak}},\
  }\href@noop {} {\bibfield  {journal} {\bibinfo  {journal} {Phys. Plasmas}\ }
  (\bibinfo {year} {2016})},\ \bibinfo {note} {submitted,
  arXiv:1607.03857}\BibitemShut {NoStop}%
\bibitem [{\citenamefont {Dahlin}, \citenamefont {Drake},\ and\ \citenamefont
  {Swisdak}(2014)}]{dahlin14a}%
  \BibitemOpen
  \bibfield  {author} {\bibinfo {author} {\bibfnamefont {J.~T.}\ \bibnamefont
  {Dahlin}}, \bibinfo {author} {\bibfnamefont {J.~F.}\ \bibnamefont {Drake}}, \
  and\ \bibinfo {author} {\bibfnamefont {M.}~\bibnamefont {Swisdak}},\ }\href
  {\doibase http://dx.doi.org/10.1063/1.4894484} {\bibfield  {journal}
  {\bibinfo  {journal} {Phys. Plasmas}\ }\textbf {\bibinfo {volume} {21}},\
  \bibinfo {eid} {092304} (\bibinfo {year} {2014})}\BibitemShut {NoStop}%
\bibitem [{\citenamefont {Wang}\ \emph {et~al.}(2016)\citenamefont {Wang},
  \citenamefont {Lu}, \citenamefont {Huang},\ and\ \citenamefont
  {Wang}}]{wang16a}%
  \BibitemOpen
  \bibfield  {author} {\bibinfo {author} {\bibfnamefont {H.}~\bibnamefont
  {Wang}}, \bibinfo {author} {\bibfnamefont {Q.}~\bibnamefont {Lu}}, \bibinfo
  {author} {\bibfnamefont {C.}~\bibnamefont {Huang}}, \ and\ \bibinfo {author}
  {\bibfnamefont {S.}~\bibnamefont {Wang}},\ }\href
  {http://stacks.iop.org/0004-637X/821/i=2/a=84} {\bibfield  {journal}
  {\bibinfo  {journal} {The Astrophysical Journal}\ }\textbf {\bibinfo {volume}
  {821}},\ \bibinfo {pages} {84} (\bibinfo {year} {2016})}\BibitemShut
  {NoStop}%
\bibitem [{\citenamefont {Guo}\ \emph {et~al.}(2014)\citenamefont {Guo},
  \citenamefont {Li}, \citenamefont {Daughton},\ and\ \citenamefont
  {Liu}}]{guo14a}%
  \BibitemOpen
  \bibfield  {author} {\bibinfo {author} {\bibfnamefont {F.}~\bibnamefont
  {Guo}}, \bibinfo {author} {\bibfnamefont {H.}~\bibnamefont {Li}}, \bibinfo
  {author} {\bibfnamefont {W.}~\bibnamefont {Daughton}}, \ and\ \bibinfo
  {author} {\bibfnamefont {Y.-H.}\ \bibnamefont {Liu}},\ }\href {\doibase
  10.1103/PhysRevLett.113.155005} {\bibfield  {journal} {\bibinfo  {journal}
  {Phys. Rev. Lett.}\ }\textbf {\bibinfo {volume} {113}},\ \bibinfo {pages}
  {155005} (\bibinfo {year} {2014})}\BibitemShut {NoStop}%
\bibitem [{\citenamefont {Li}\ \emph {et~al.}(2015)\citenamefont {Li},
  \citenamefont {Guo}, \citenamefont {Li},\ and\ \citenamefont {Li}}]{li15a}%
  \BibitemOpen
  \bibfield  {author} {\bibinfo {author} {\bibfnamefont {X.}~\bibnamefont
  {Li}}, \bibinfo {author} {\bibfnamefont {F.}~\bibnamefont {Guo}}, \bibinfo
  {author} {\bibfnamefont {H.}~\bibnamefont {Li}}, \ and\ \bibinfo {author}
  {\bibfnamefont {G.}~\bibnamefont {Li}},\ }\href
  {http://stacks.iop.org/2041-8205/811/i=2/a=L24} {\bibfield  {journal}
  {\bibinfo  {journal} {The Astrophysical Journal Letters}\ }\textbf {\bibinfo
  {volume} {811}},\ \bibinfo {pages} {L24} (\bibinfo {year}
  {2015})}\BibitemShut {NoStop}%
\bibitem [{\citenamefont {Numata}\ and\ \citenamefont
  {Loureiro}(2015)}]{numata15a}%
  \BibitemOpen
  \bibfield  {author} {\bibinfo {author} {\bibfnamefont {R.}~\bibnamefont
  {Numata}}\ and\ \bibinfo {author} {\bibfnamefont {N.~F.}\ \bibnamefont
  {Loureiro}},\ }\href@noop {} {\bibfield  {journal} {\bibinfo  {journal}
  {Journal of Plasma Physics}\ }\textbf {\bibinfo {volume} {81}},\ \bibinfo
  {pages} {305810201 (17 pages)} (\bibinfo {year} {2015})}\BibitemShut
  {NoStop}%
\bibitem [{\citenamefont {Chen}\ \emph {et~al.}(2008)\citenamefont {Chen},
  \citenamefont {Bhattacharjee}, \citenamefont {Puhl-Quinn}, \citenamefont
  {Yang}, \citenamefont {Bessho}, \citenamefont {Imada}, \citenamefont
  {M\"{u}hlbachler}, \citenamefont {Daly}, \citenamefont {Lefebvre},
  \citenamefont {Khotyaintsev}, \citenamefont {Vaivads}, \citenamefont
  {Fazakerley},\ and\ \citenamefont {Georgescu}}]{chen08a}%
  \BibitemOpen
  \bibfield  {author} {\bibinfo {author} {\bibfnamefont {L.-J.}\ \bibnamefont
  {Chen}}, \bibinfo {author} {\bibfnamefont {A.}~\bibnamefont {Bhattacharjee}},
  \bibinfo {author} {\bibfnamefont {P.~A.}\ \bibnamefont {Puhl-Quinn}},
  \bibinfo {author} {\bibfnamefont {H.}~\bibnamefont {Yang}}, \bibinfo {author}
  {\bibfnamefont {N.}~\bibnamefont {Bessho}}, \bibinfo {author} {\bibfnamefont
  {S.}~\bibnamefont {Imada}}, \bibinfo {author} {\bibfnamefont
  {S.}~\bibnamefont {M\"{u}hlbachler}}, \bibinfo {author} {\bibfnamefont
  {P.~W.}\ \bibnamefont {Daly}}, \bibinfo {author} {\bibfnamefont
  {B.}~\bibnamefont {Lefebvre}}, \bibinfo {author} {\bibfnamefont
  {Y.}~\bibnamefont {Khotyaintsev}}, \bibinfo {author} {\bibfnamefont
  {A.}~\bibnamefont {Vaivads}}, \bibinfo {author} {\bibfnamefont
  {A.}~\bibnamefont {Fazakerley}}, \ and\ \bibinfo {author} {\bibfnamefont
  {E.}~\bibnamefont {Georgescu}},\ }\href {\doibase 10.1038/nphys777}
  {\bibfield  {journal} {\bibinfo  {journal} {Nature Physics}\ }\textbf
  {\bibinfo {volume} {4}},\ \bibinfo {pages} {19} (\bibinfo {year}
  {2008})}\BibitemShut {NoStop}%
\bibitem [{\citenamefont {Retin{\`o}}\ \emph {et~al.}(2008)\citenamefont
  {Retin{\`o}}, \citenamefont {Nakamura}, \citenamefont {Vaivads},
  \citenamefont {Khotyaintsev}, \citenamefont {Hayakawa}, \citenamefont
  {Tanaka}, \citenamefont {Kasahara}, \citenamefont {Fujimoto}, \citenamefont
  {Shinohara}, \citenamefont {Eastwood}, \citenamefont {André}, \citenamefont
  {Baumjohann}, \citenamefont {Daly}, \citenamefont {Kronberg},\ and\
  \citenamefont {Cornilleau-Wehrlin}}]{retino08a}%
  \BibitemOpen
  \bibfield  {author} {\bibinfo {author} {\bibfnamefont {A.}~\bibnamefont
  {Retin{\`o}}}, \bibinfo {author} {\bibfnamefont {R.}~\bibnamefont
  {Nakamura}}, \bibinfo {author} {\bibfnamefont {A.}~\bibnamefont {Vaivads}},
  \bibinfo {author} {\bibfnamefont {Y.}~\bibnamefont {Khotyaintsev}}, \bibinfo
  {author} {\bibfnamefont {T.}~\bibnamefont {Hayakawa}}, \bibinfo {author}
  {\bibfnamefont {K.}~\bibnamefont {Tanaka}}, \bibinfo {author} {\bibfnamefont
  {S.}~\bibnamefont {Kasahara}}, \bibinfo {author} {\bibfnamefont
  {M.}~\bibnamefont {Fujimoto}}, \bibinfo {author} {\bibfnamefont
  {I.}~\bibnamefont {Shinohara}}, \bibinfo {author} {\bibfnamefont {J.~P.}\
  \bibnamefont {Eastwood}}, \bibinfo {author} {\bibfnamefont {M.}~\bibnamefont
  {André}}, \bibinfo {author} {\bibfnamefont {W.}~\bibnamefont {Baumjohann}},
  \bibinfo {author} {\bibfnamefont {P.~W.}\ \bibnamefont {Daly}}, \bibinfo
  {author} {\bibfnamefont {E.~A.}\ \bibnamefont {Kronberg}}, \ and\ \bibinfo
  {author} {\bibfnamefont {N.}~\bibnamefont {Cornilleau-Wehrlin}},\ }\href
  {\doibase 10.1029/2008JA013511} {\bibfield  {journal} {\bibinfo  {journal}
  {Journal of Geophysical Research: Space Physics}\ }\textbf {\bibinfo {volume}
  {113}} (\bibinfo {year} {2008}),\ 10.1029/2008JA013511},\ \bibinfo {note}
  {a12215}\BibitemShut {NoStop}%
\bibitem [{\citenamefont {Huang}\ \emph {et~al.}(2012)\citenamefont {Huang},
  \citenamefont {Vaivads}, \citenamefont {Khotyaintsev}, \citenamefont {Zhou},
  \citenamefont {Fu}, \citenamefont {Retinò}, \citenamefont {Deng},
  \citenamefont {André}, \citenamefont {Cully}, \citenamefont {He},
  \citenamefont {Sahraoui}, \citenamefont {Yuan},\ and\ \citenamefont
  {Pang}}]{huang12a}%
  \BibitemOpen
  \bibfield  {author} {\bibinfo {author} {\bibfnamefont {S.~Y.}\ \bibnamefont
  {Huang}}, \bibinfo {author} {\bibfnamefont {A.}~\bibnamefont {Vaivads}},
  \bibinfo {author} {\bibfnamefont {Y.~V.}\ \bibnamefont {Khotyaintsev}},
  \bibinfo {author} {\bibfnamefont {M.}~\bibnamefont {Zhou}}, \bibinfo {author}
  {\bibfnamefont {H.~S.}\ \bibnamefont {Fu}}, \bibinfo {author} {\bibfnamefont
  {A.}~\bibnamefont {Retinò}}, \bibinfo {author} {\bibfnamefont {X.~H.}\
  \bibnamefont {Deng}}, \bibinfo {author} {\bibfnamefont {M.}~\bibnamefont
  {André}}, \bibinfo {author} {\bibfnamefont {C.~M.}\ \bibnamefont {Cully}},
  \bibinfo {author} {\bibfnamefont {J.~S.}\ \bibnamefont {He}}, \bibinfo
  {author} {\bibfnamefont {F.}~\bibnamefont {Sahraoui}}, \bibinfo {author}
  {\bibfnamefont {Z.~G.}\ \bibnamefont {Yuan}}, \ and\ \bibinfo {author}
  {\bibfnamefont {Y.}~\bibnamefont {Pang}},\ }\href {\doibase
  10.1029/2012GL051946} {\bibfield  {journal} {\bibinfo  {journal} {Geophysical
  Research Letters}\ }\textbf {\bibinfo {volume} {39}} (\bibinfo {year}
  {2012}),\ 10.1029/2012GL051946},\ \bibinfo {note} {l11103}\BibitemShut
  {NoStop}%
\bibitem [{\citenamefont {Schreier}\ \emph {et~al.}(2010)\citenamefont
  {Schreier}, \citenamefont {Swisdak}, \citenamefont {Drake},\ and\
  \citenamefont {Cassak}}]{schreier10a}%
  \BibitemOpen
  \bibfield  {author} {\bibinfo {author} {\bibfnamefont {R.}~\bibnamefont
  {Schreier}}, \bibinfo {author} {\bibfnamefont {M.}~\bibnamefont {Swisdak}},
  \bibinfo {author} {\bibfnamefont {J.~F.}\ \bibnamefont {Drake}}, \ and\
  \bibinfo {author} {\bibfnamefont {P.~A.}\ \bibnamefont {Cassak}},\ }\href
  {\doibase 10.1063/1.3494218} {\bibfield  {journal} {\bibinfo  {journal}
  {Phys. Plasmas}\ }\textbf {\bibinfo {volume} {17}},\ \bibinfo {eid} {110704}
  (\bibinfo {year} {2010}),\ 10.1063/1.3494218}\BibitemShut {NoStop}%
\bibitem [{\citenamefont {Daughton}\ \emph {et~al.}(2011)\citenamefont
  {Daughton}, \citenamefont {Roytershteyn}, \citenamefont {Karimabadi},
  \citenamefont {Yin}, \citenamefont {Albright}, \citenamefont {Bergen},\ and\
  \citenamefont {Bowers}}]{daughton11a}%
  \BibitemOpen
  \bibfield  {author} {\bibinfo {author} {\bibfnamefont {W.}~\bibnamefont
  {Daughton}}, \bibinfo {author} {\bibfnamefont {V.}~\bibnamefont
  {Roytershteyn}}, \bibinfo {author} {\bibfnamefont {H.}~\bibnamefont
  {Karimabadi}}, \bibinfo {author} {\bibfnamefont {L.}~\bibnamefont {Yin}},
  \bibinfo {author} {\bibfnamefont {B.~J.}\ \bibnamefont {Albright}}, \bibinfo
  {author} {\bibfnamefont {B.}~\bibnamefont {Bergen}}, \ and\ \bibinfo {author}
  {\bibfnamefont {K.~J.}\ \bibnamefont {Bowers}},\ }\href {\doibase
  10.1038/nphys1965} {\bibfield  {journal} {\bibinfo  {journal} {Nature Phys.}\
  }\textbf {\bibinfo {volume} {7}},\ \bibinfo {pages} {539} (\bibinfo {year}
  {2011})}\BibitemShut {NoStop}%
\bibitem [{\citenamefont {Onofri}, \citenamefont {Isliker},\ and\ \citenamefont
  {Vlahos}(2006)}]{onofri06a}%
  \BibitemOpen
  \bibfield  {author} {\bibinfo {author} {\bibfnamefont {M.}~\bibnamefont
  {Onofri}}, \bibinfo {author} {\bibfnamefont {H.}~\bibnamefont {Isliker}}, \
  and\ \bibinfo {author} {\bibfnamefont {L.}~\bibnamefont {Vlahos}},\ }\href
  {\doibase 10.1103/PhysRevLett96.151102} {\bibfield  {journal} {\bibinfo
  {journal} {Phys. Rev. Lett.}\ }\textbf {\bibinfo {volume} {96}},\ \bibinfo
  {eid} {151102} (\bibinfo {year} {2006}),\
  10.1103/PhysRevLett96.151102}\BibitemShut {NoStop}%
\bibitem [{\citenamefont {Kowal}, \citenamefont {de~Gouveia Dal~Pino},\ and\
  \citenamefont {Lazarian}(2011)}]{kowal11a}%
  \BibitemOpen
  \bibfield  {author} {\bibinfo {author} {\bibfnamefont {G.}~\bibnamefont
  {Kowal}}, \bibinfo {author} {\bibfnamefont {E.~M.}\ \bibnamefont {de~Gouveia
  Dal~Pino}}, \ and\ \bibinfo {author} {\bibfnamefont {A.}~\bibnamefont
  {Lazarian}},\ }\href {http://stacks.iop.org/0004-637X/735/i=2/a=102}
  {\bibfield  {journal} {\bibinfo  {journal} {The Astrophysical Journal}\
  }\textbf {\bibinfo {volume} {735}},\ \bibinfo {pages} {102} (\bibinfo {year}
  {2011})}\BibitemShut {NoStop}%
\bibitem [{\citenamefont {Sironi}\ and\ \citenamefont
  {Spitkovsky}(2014)}]{sironi14a}%
  \BibitemOpen
  \bibfield  {author} {\bibinfo {author} {\bibfnamefont {L.}~\bibnamefont
  {Sironi}}\ and\ \bibinfo {author} {\bibfnamefont {A.}~\bibnamefont
  {Spitkovsky}},\ }\href {http://stacks.iop.org/2041-8205/783/i=1/a=L21}
  {\bibfield  {journal} {\bibinfo  {journal} {The Astrophysical Journal
  Letters}\ }\textbf {\bibinfo {volume} {783}},\ \bibinfo {pages} {L21}
  (\bibinfo {year} {2014})}\BibitemShut {NoStop}%
\bibitem [{\citenamefont {Northrop}(1963)}]{northrop63a}%
  \BibitemOpen
  \bibfield  {author} {\bibinfo {author} {\bibfnamefont {T.~G.}\ \bibnamefont
  {Northrop}},\ }\href {\doibase 10.1029/RG001i003p00283} {\bibfield  {journal}
  {\bibinfo  {journal} {Reviews of Geophysics}\ }\textbf {\bibinfo {volume}
  {1}},\ \bibinfo {pages} {283} (\bibinfo {year} {1963})}\BibitemShut {NoStop}%
\bibitem [{\citenamefont {Zeiler}\ \emph {et~al.}(2002)\citenamefont {Zeiler},
  \citenamefont {Biskamp}, \citenamefont {Drake}, \citenamefont {Rogers},
  \citenamefont {Shay},\ and\ \citenamefont {Scholer}}]{zeiler02a}%
  \BibitemOpen
  \bibfield  {author} {\bibinfo {author} {\bibfnamefont {A.}~\bibnamefont
  {Zeiler}}, \bibinfo {author} {\bibfnamefont {D.}~\bibnamefont {Biskamp}},
  \bibinfo {author} {\bibfnamefont {J.~F.}\ \bibnamefont {Drake}}, \bibinfo
  {author} {\bibfnamefont {B.~N.}\ \bibnamefont {Rogers}}, \bibinfo {author}
  {\bibfnamefont {M.~A.}\ \bibnamefont {Shay}}, \ and\ \bibinfo {author}
  {\bibfnamefont {M.}~\bibnamefont {Scholer}},\ }\href {\doibase
  10.1029/2001JA000287} {\bibfield  {journal} {\bibinfo  {journal} {J. Geophys.
  Res.}\ }\textbf {\bibinfo {volume} {107}},\ \bibinfo {pages} {1230} (\bibinfo
  {year} {2002})}\BibitemShut {NoStop}%
\bibitem [{\citenamefont {Bobrova}\ \emph {et~al.}(2001)\citenamefont
  {Bobrova}, \citenamefont {Bulanov}, \citenamefont {Sakai},\ and\
  \citenamefont {Sugiyama}}]{bobrova01a}%
  \BibitemOpen
  \bibfield  {author} {\bibinfo {author} {\bibfnamefont {N.~A.}\ \bibnamefont
  {Bobrova}}, \bibinfo {author} {\bibfnamefont {S.~V.}\ \bibnamefont
  {Bulanov}}, \bibinfo {author} {\bibfnamefont {J.~I.}\ \bibnamefont {Sakai}},
  \ and\ \bibinfo {author} {\bibfnamefont {D.}~\bibnamefont {Sugiyama}},\
  }\href@noop {} {\bibfield  {journal} {\bibinfo  {journal} {Physics of
  Plasmas}\ }\textbf {\bibinfo {volume} {8}} (\bibinfo {year}
  {2001})}\BibitemShut {NoStop}%
\bibitem [{\citenamefont {Furth}, \citenamefont {Killeen},\ and\ \citenamefont
  {Rosenbluth}(1963)}]{furth63a}%
  \BibitemOpen
  \bibfield  {author} {\bibinfo {author} {\bibfnamefont {H.~P.}\ \bibnamefont
  {Furth}}, \bibinfo {author} {\bibfnamefont {J.}~\bibnamefont {Killeen}}, \
  and\ \bibinfo {author} {\bibfnamefont {M.~N.}\ \bibnamefont {Rosenbluth}},\
  }\href@noop {} {\bibfield  {journal} {\bibinfo  {journal} {Physics of
  Fluids}\ }\textbf {\bibinfo {volume} {6}} (\bibinfo {year}
  {1963})}\BibitemShut {NoStop}%
\bibitem [{\citenamefont {Rechester}\ and\ \citenamefont
  {Rosenbluth}(1978)}]{rechester78a}%
  \BibitemOpen
  \bibfield  {author} {\bibinfo {author} {\bibfnamefont {A.~B.}\ \bibnamefont
  {Rechester}}\ and\ \bibinfo {author} {\bibfnamefont {M.~N.}\ \bibnamefont
  {Rosenbluth}},\ }\href {\doibase 10.1103/PhysRevLett.40.38} {\bibfield
  {journal} {\bibinfo  {journal} {Phys. Rev. Lett.}\ }\textbf {\bibinfo
  {volume} {40}},\ \bibinfo {pages} {38} (\bibinfo {year} {1978})}\BibitemShut
  {NoStop}%
\bibitem [{\citenamefont {Daughton}\ \emph {et~al.}(2014)\citenamefont
  {Daughton}, \citenamefont {Nakamura}, \citenamefont {Karimabadi},
  \citenamefont {Roytershteyn},\ and\ \citenamefont {Loring}}]{daughton14a}%
  \BibitemOpen
  \bibfield  {author} {\bibinfo {author} {\bibfnamefont {W.}~\bibnamefont
  {Daughton}}, \bibinfo {author} {\bibfnamefont {T.~K.~M.}\ \bibnamefont
  {Nakamura}}, \bibinfo {author} {\bibfnamefont {H.}~\bibnamefont
  {Karimabadi}}, \bibinfo {author} {\bibfnamefont {V.}~\bibnamefont
  {Roytershteyn}}, \ and\ \bibinfo {author} {\bibfnamefont {B.}~\bibnamefont
  {Loring}},\ }\href {\doibase http://dx.doi.org/10.1063/1.4875730} {\bibfield
  {journal} {\bibinfo  {journal} {Physics of Plasmas}\ }\textbf {\bibinfo
  {volume} {21}},\ \bibinfo {eid} {052307} (\bibinfo {year} {2014}),\
  http://dx.doi.org/10.1063/1.4875730}\BibitemShut {NoStop}%
\bibitem [{\citenamefont {Schoeffler}, \citenamefont {Drake},\ and\
  \citenamefont {Swisdak}(2011)}]{schoeffler11a}%
  \BibitemOpen
  \bibfield  {author} {\bibinfo {author} {\bibfnamefont {K.~M.}\ \bibnamefont
  {Schoeffler}}, \bibinfo {author} {\bibfnamefont {J.~F.}\ \bibnamefont
  {Drake}}, \ and\ \bibinfo {author} {\bibfnamefont {M.}~\bibnamefont
  {Swisdak}},\ }\href {\doibase 10.1088/0004-637X/743/1/70} {\bibfield
  {journal} {\bibinfo  {journal} {Ap. J.}\ }\textbf {\bibinfo {volume} {743}},\
  \bibinfo {pages} {70} (\bibinfo {year} {2011})}\BibitemShut {NoStop}%
\bibitem [{\citenamefont {Kasper}, \citenamefont {Lazarus},\ and\ \citenamefont
  {Gary}(2002)}]{kasper02a}%
  \BibitemOpen
  \bibfield  {author} {\bibinfo {author} {\bibfnamefont {J.~C.}\ \bibnamefont
  {Kasper}}, \bibinfo {author} {\bibfnamefont {A.~J.}\ \bibnamefont {Lazarus}},
  \ and\ \bibinfo {author} {\bibfnamefont {S.~P.}\ \bibnamefont {Gary}},\
  }\href {\doibase 10.1029/2002GL015128} {\bibfield  {journal} {\bibinfo
  {journal} {Geophysical Research Letters}\ }\textbf {\bibinfo {volume} {29}},\
  \bibinfo {pages} {20} (\bibinfo {year} {2002})},\ \bibinfo {note}
  {1839}\BibitemShut {NoStop}%
\bibitem [{\citenamefont {Bale}\ \emph {et~al.}(2009)\citenamefont {Bale},
  \citenamefont {Kasper}, \citenamefont {Howes}, \citenamefont {Quataert},
  \citenamefont {Salem},\ and\ \citenamefont {Sundkvist}}]{bale09a}%
  \BibitemOpen
  \bibfield  {author} {\bibinfo {author} {\bibfnamefont {S.~D.}\ \bibnamefont
  {Bale}}, \bibinfo {author} {\bibfnamefont {J.~C.}\ \bibnamefont {Kasper}},
  \bibinfo {author} {\bibfnamefont {G.~G.}\ \bibnamefont {Howes}}, \bibinfo
  {author} {\bibfnamefont {E.}~\bibnamefont {Quataert}}, \bibinfo {author}
  {\bibfnamefont {C.}~\bibnamefont {Salem}}, \ and\ \bibinfo {author}
  {\bibfnamefont {D.}~\bibnamefont {Sundkvist}},\ }\href {\doibase
  10.1103/PhysRevLett.103.211101} {\bibfield  {journal} {\bibinfo  {journal}
  {Phys. Rev. Lett.}\ }\textbf {\bibinfo {volume} {103}},\ \bibinfo {eid}
  {211101} (\bibinfo {year} {2009}),\
  10.1103/PhysRevLett.103.211101}\BibitemShut {NoStop}%
\bibitem [{\citenamefont {{Jokipii}}\ and\ \citenamefont
  {{Parker}}(1969)}]{jokipii69a}%
  \BibitemOpen
  \bibfield  {author} {\bibinfo {author} {\bibfnamefont {J.~R.}\ \bibnamefont
  {{Jokipii}}}\ and\ \bibinfo {author} {\bibfnamefont {E.~N.}\ \bibnamefont
  {{Parker}}},\ }\href {\doibase 10.1086/149909} {\bibfield  {journal}
  {\bibinfo  {journal} {The Astrophysical Journal}\ }\textbf {\bibinfo {volume}
  {155}},\ \bibinfo {pages} {777} (\bibinfo {year} {1969})}\BibitemShut
  {NoStop}%
\bibitem [{\citenamefont {Shay}\ \emph {et~al.}(2014)\citenamefont {Shay},
  \citenamefont {Haggerty}, \citenamefont {Phan}, \citenamefont {Drake},
  \citenamefont {Cassak}, \citenamefont {Wu}, \citenamefont {Oieroset},
  \citenamefont {Swisdak},\ and\ \citenamefont {Malakit}}]{shay14a}%
  \BibitemOpen
  \bibfield  {author} {\bibinfo {author} {\bibfnamefont {M.~A.}\ \bibnamefont
  {Shay}}, \bibinfo {author} {\bibfnamefont {C.~C.}\ \bibnamefont {Haggerty}},
  \bibinfo {author} {\bibfnamefont {T.~D.}\ \bibnamefont {Phan}}, \bibinfo
  {author} {\bibfnamefont {J.~F.}\ \bibnamefont {Drake}}, \bibinfo {author}
  {\bibfnamefont {P.~A.}\ \bibnamefont {Cassak}}, \bibinfo {author}
  {\bibfnamefont {P.}~\bibnamefont {Wu}}, \bibinfo {author} {\bibfnamefont
  {M.}~\bibnamefont {Oieroset}}, \bibinfo {author} {\bibfnamefont
  {M.}~\bibnamefont {Swisdak}}, \ and\ \bibinfo {author} {\bibfnamefont
  {K.}~\bibnamefont {Malakit}},\ }\href {\doibase
  http://dx.doi.org/10.1063/1.4904203} {\bibfield  {journal} {\bibinfo
  {journal} {Phys. Plasmas}\ }\textbf {\bibinfo {volume} {21}},\ \bibinfo {eid}
  {122902} (\bibinfo {year} {2014})}\BibitemShut {NoStop}%
\bibitem [{\citenamefont {Haggerty}\ \emph {et~al.}(2015)\citenamefont
  {Haggerty}, \citenamefont {Shay}, \citenamefont {Drake}, \citenamefont
  {Phan},\ and\ \citenamefont {McHugh}}]{haggerty15a}%
  \BibitemOpen
  \bibfield  {author} {\bibinfo {author} {\bibfnamefont {C.~C.}\ \bibnamefont
  {Haggerty}}, \bibinfo {author} {\bibfnamefont {M.~A.}\ \bibnamefont {Shay}},
  \bibinfo {author} {\bibfnamefont {J.~F.}\ \bibnamefont {Drake}}, \bibinfo
  {author} {\bibfnamefont {T.~D.}\ \bibnamefont {Phan}}, \ and\ \bibinfo
  {author} {\bibfnamefont {C.~T.}\ \bibnamefont {McHugh}},\ }\href {\doibase
  10.1002/2015GL065961} {\bibfield  {journal} {\bibinfo  {journal} {Geophysical
  Research Letters}\ }\textbf {\bibinfo {volume} {42}},\ \bibinfo {pages}
  {9657} (\bibinfo {year} {2015})},\ \bibinfo {note} {2015GL065961}\BibitemShut
  {NoStop}%
\bibitem [{\citenamefont {Drake}\ \emph
  {et~al.}(2009{\natexlab{a}})\citenamefont {Drake}, \citenamefont {Swisdak},
  \citenamefont {Phan}, \citenamefont {Cassak}, \citenamefont {Shay},
  \citenamefont {Lepri}, \citenamefont {Lon}, \citenamefont {Quataert},\ and\
  \citenamefont {Zurbuchen}}]{drake09a}%
  \BibitemOpen
  \bibfield  {author} {\bibinfo {author} {\bibfnamefont {J.~F.}\ \bibnamefont
  {Drake}}, \bibinfo {author} {\bibfnamefont {M.}~\bibnamefont {Swisdak}},
  \bibinfo {author} {\bibfnamefont {T.~D.}\ \bibnamefont {Phan}}, \bibinfo
  {author} {\bibfnamefont {P.~A.}\ \bibnamefont {Cassak}}, \bibinfo {author}
  {\bibfnamefont {M.~A.}\ \bibnamefont {Shay}}, \bibinfo {author}
  {\bibfnamefont {S.~T.}\ \bibnamefont {Lepri}}, \bibinfo {author}
  {\bibfnamefont {R.~P.}\ \bibnamefont {Lon}}, \bibinfo {author} {\bibfnamefont
  {E.}~\bibnamefont {Quataert}}, \ and\ \bibinfo {author} {\bibfnamefont
  {T.~H.}\ \bibnamefont {Zurbuchen}},\ }\href {\doibase 10.1029/2008JA013701}
  {\bibfield  {journal} {\bibinfo  {journal} {J. Geophys. Res.}\ }\textbf
  {\bibinfo {volume} {114}},\ \bibinfo {eid} {A05111} (\bibinfo {year}
  {2009}{\natexlab{a}}),\ 10.1029/2008JA013701}\BibitemShut {NoStop}%
\bibitem [{\citenamefont {Drake}\ \emph
  {et~al.}(2009{\natexlab{b}})\citenamefont {Drake}, \citenamefont {Cassak},
  \citenamefont {Shay}, \citenamefont {Swisdak},\ and\ \citenamefont
  {Quataert}}]{drake09b}%
  \BibitemOpen
  \bibfield  {author} {\bibinfo {author} {\bibfnamefont {J.~F.}\ \bibnamefont
  {Drake}}, \bibinfo {author} {\bibfnamefont {P.~A.}\ \bibnamefont {Cassak}},
  \bibinfo {author} {\bibfnamefont {M.~A.}\ \bibnamefont {Shay}}, \bibinfo
  {author} {\bibfnamefont {M.}~\bibnamefont {Swisdak}}, \ and\ \bibinfo
  {author} {\bibfnamefont {E.}~\bibnamefont {Quataert}},\ }\href {\doibase
  10.1088/0004-637X/700/1/L16} {\bibfield  {journal} {\bibinfo  {journal} {Ap.
  J.}\ }\textbf {\bibinfo {volume} {700}},\ \bibinfo {pages} {L16} (\bibinfo
  {year} {2009}{\natexlab{b}})}\BibitemShut {NoStop}%
\bibitem [{\citenamefont {Knizhnik}, \citenamefont {Swisdak},\ and\
  \citenamefont {Drake}(2011)}]{knizhnik11a}%
  \BibitemOpen
  \bibfield  {author} {\bibinfo {author} {\bibfnamefont {K.}~\bibnamefont
  {Knizhnik}}, \bibinfo {author} {\bibfnamefont {M.}~\bibnamefont {Swisdak}}, \
  and\ \bibinfo {author} {\bibfnamefont {J.}~\bibnamefont {Drake}},\ }\href
  {\doibase 10.1088/2041-8205/743/2/L35} {\bibfield  {journal} {\bibinfo
  {journal} {Ap. J. Lett.}\ }\textbf {\bibinfo {volume} {743}} (\bibinfo {year}
  {2011}),\ 10.1088/2041-8205/743/2/L35}\BibitemShut {NoStop}%
\bibitem [{\citenamefont {Guo}\ \emph {et~al.}(2015)\citenamefont {Guo},
  \citenamefont {Liu}, \citenamefont {Daughton},\ and\ \citenamefont
  {Li}}]{guo15a}%
  \BibitemOpen
  \bibfield  {author} {\bibinfo {author} {\bibfnamefont {F.}~\bibnamefont
  {Guo}}, \bibinfo {author} {\bibfnamefont {Y.-H.}\ \bibnamefont {Liu}},
  \bibinfo {author} {\bibfnamefont {W.}~\bibnamefont {Daughton}}, \ and\
  \bibinfo {author} {\bibfnamefont {H.}~\bibnamefont {Li}},\ }\href
  {http://stacks.iop.org/0004-637X/806/i=2/a=167} {\bibfield  {journal}
  {\bibinfo  {journal} {The Astrophysical Journal}\ }\textbf {\bibinfo {volume}
  {806}},\ \bibinfo {pages} {167} (\bibinfo {year} {2015})}\BibitemShut
  {NoStop}%
\bibitem [{\citenamefont {Guo}\ \emph {et~al.}(2016)\citenamefont {Guo},
  \citenamefont {Li}, \citenamefont {Daughton}, \citenamefont {Li},\ and\
  \citenamefont {Liu}}]{guo16a}%
  \BibitemOpen
  \bibfield  {author} {\bibinfo {author} {\bibfnamefont {F.}~\bibnamefont
  {Guo}}, \bibinfo {author} {\bibfnamefont {H.}~\bibnamefont {Li}}, \bibinfo
  {author} {\bibfnamefont {W.}~\bibnamefont {Daughton}}, \bibinfo {author}
  {\bibfnamefont {X.}~\bibnamefont {Li}}, \ and\ \bibinfo {author}
  {\bibfnamefont {Y.-H.}\ \bibnamefont {Liu}},\ }\href {\doibase
  http://dx.doi.org/10.1063/1.4948284} {\bibfield  {journal} {\bibinfo
  {journal} {Physics of Plasmas}\ }\textbf {\bibinfo {volume} {23}},\ \bibinfo
  {eid} {055708} (\bibinfo {year} {2016}),\
  http://dx.doi.org/10.1063/1.4948284}\BibitemShut {NoStop}%
\bibitem [{\citenamefont {Li}, \citenamefont {Drake},\ and\ \citenamefont
  {Swisdak}(2012)}]{li12a}%
  \BibitemOpen
  \bibfield  {author} {\bibinfo {author} {\bibfnamefont {T.~C.}\ \bibnamefont
  {Li}}, \bibinfo {author} {\bibfnamefont {J.~F.}\ \bibnamefont {Drake}}, \
  and\ \bibinfo {author} {\bibfnamefont {M.}~\bibnamefont {Swisdak}},\ }\href
  {http://stacks.iop.org/0004-637X/757/i=1/a=20} {\bibfield  {journal}
  {\bibinfo  {journal} {The Astrophysical Journal}\ }\textbf {\bibinfo {volume}
  {757}},\ \bibinfo {pages} {20} (\bibinfo {year} {2012})}\BibitemShut
  {NoStop}%
\bibitem [{\citenamefont {Werner}\ \emph {et~al.}(2016)\citenamefont {Werner},
  \citenamefont {Uzdensky}, \citenamefont {Cerutti}, \citenamefont
  {Nalewajko},\ and\ \citenamefont {Begelman}}]{werner15a}%
  \BibitemOpen
  \bibfield  {author} {\bibinfo {author} {\bibfnamefont {G.~R.}\ \bibnamefont
  {Werner}}, \bibinfo {author} {\bibfnamefont {D.~A.}\ \bibnamefont
  {Uzdensky}}, \bibinfo {author} {\bibfnamefont {B.}~\bibnamefont {Cerutti}},
  \bibinfo {author} {\bibfnamefont {K.}~\bibnamefont {Nalewajko}}, \ and\
  \bibinfo {author} {\bibfnamefont {M.~C.}\ \bibnamefont {Begelman}},\ }\href
  {http://stacks.iop.org/2041-8205/816/i=1/a=L8} {\bibfield  {journal}
  {\bibinfo  {journal} {The Astrophysical Journal Letters}\ }\textbf {\bibinfo
  {volume} {816}},\ \bibinfo {pages} {L8} (\bibinfo {year} {2016})}\BibitemShut
  {NoStop}%
\bibitem [{\citenamefont {Fletcher}\ \emph {et~al.}(2011)\citenamefont
  {Fletcher}, \citenamefont {Dennis}, \citenamefont {Hudson}, \citenamefont
  {Krucker}, \citenamefont {Phillips}, \citenamefont {Veronig}, \citenamefont
  {Battaglia}, \citenamefont {Bone}, \citenamefont {Caspi}, \citenamefont
  {Chen}, \citenamefont {Gallagher}, \citenamefont {Grigis}, \citenamefont
  {Ji}, \citenamefont {Liu}, \citenamefont {Milligan},\ and\ \citenamefont
  {Temmer}}]{fletcher11a}%
  \BibitemOpen
  \bibfield  {author} {\bibinfo {author} {\bibfnamefont {L.}~\bibnamefont
  {Fletcher}}, \bibinfo {author} {\bibfnamefont {B.~R.}\ \bibnamefont
  {Dennis}}, \bibinfo {author} {\bibfnamefont {H.~S.}\ \bibnamefont {Hudson}},
  \bibinfo {author} {\bibfnamefont {S.}~\bibnamefont {Krucker}}, \bibinfo
  {author} {\bibfnamefont {K.}~\bibnamefont {Phillips}}, \bibinfo {author}
  {\bibfnamefont {A.}~\bibnamefont {Veronig}}, \bibinfo {author} {\bibfnamefont
  {M.}~\bibnamefont {Battaglia}}, \bibinfo {author} {\bibfnamefont
  {L.}~\bibnamefont {Bone}}, \bibinfo {author} {\bibfnamefont {A.}~\bibnamefont
  {Caspi}}, \bibinfo {author} {\bibfnamefont {Q.}~\bibnamefont {Chen}},
  \bibinfo {author} {\bibfnamefont {P.}~\bibnamefont {Gallagher}}, \bibinfo
  {author} {\bibfnamefont {P.~T.}\ \bibnamefont {Grigis}}, \bibinfo {author}
  {\bibfnamefont {H.}~\bibnamefont {Ji}}, \bibinfo {author} {\bibfnamefont
  {W.}~\bibnamefont {Liu}}, \bibinfo {author} {\bibfnamefont {R.~O.}\
  \bibnamefont {Milligan}}, \ and\ \bibinfo {author} {\bibfnamefont
  {M.}~\bibnamefont {Temmer}},\ }\href {\doibase 10.1007/s11214-010-9701-8}
  {\bibfield  {journal} {\bibinfo  {journal} {Space Science Reviews}\ }\textbf
  {\bibinfo {volume} {159}},\ \bibinfo {pages} {19} (\bibinfo {year}
  {2011})}\BibitemShut {NoStop}%
\end{thebibliography}%

\end{document}